\documentclass[aps,%
twocolumn,%
floatfix,%
pra,%
amsfonts,%
groupedaddress,%
superscriptaddress]{revtex4-2}%
\usepackage[utf8]{inputenc}
\usepackage[T1]{fontenc}
\usepackage{mathrsfs}       
\usepackage{mathtools}
\usepackage[titletoc,title]{appendix}
\usepackage{braket}
\usepackage{verbatim}
\usepackage{amsmath}
\usepackage{amsthm}
\newtheorem{theorem}{Theorem}

\usepackage{booktabs}
\usepackage[linesnumbered,ruled,vlined]{algorithm2e}
\usepackage{amsmath, amssymb}
\usepackage{bm}
\usepackage{amsthm}
\newtheorem{definition}{Definition}
\usepackage{tikz}
\usetikzlibrary{shapes, arrows.meta, positioning}
\usepackage{graphicx}
\usepackage[dvipsnames]{xcolor}
\definecolor{beamer@blendedblue}{rgb}{0.2,0.2,0.7}
\usepackage[bookmarks=false]{hyperref}
\hypersetup{colorlinks=true,%
citecolor=beamer@blendedblue,%
linkcolor=beamer@blendedblue,%
urlcolor=beamer@blendedblue,%
pdfstartview=FitH,%
bookmarksopen=true}
\usepackage{makecell}
\usepackage{array}

\usepackage[osf,sc]{mathpazo}
\usepackage{dsfont}

\definecolor{googleblue}{HTML}{4285F4}
\definecolor{googlered}{HTML}{DB4437}
\definecolor{googleyellow}{HTML}{F4B400}
\definecolor{googlegreen}{HTML}{0F9D58}
\definecolor{klevinblue}{HTML}{002FA7}
\definecolor{tiffanyblue}{HTML}{0ABAB5}

\newcommand{\BipartiteGBSGA}{\texorpdfstring{\mbox{BipartiteGBS-GA}}{BipartiteGBS-GA}}
\newcommand{\MSBipartiteGBSGA}{\texorpdfstring{\mbox{MS-BipartiteGBS-GA}}{MS-BipartiteGBS-GA}}

\begin{document}

\newcommand{\thetitle}{Bipartite Gaussian Boson Sampling for Hamiltonian Cycles in Directed Graphs}
\title{\thetitle}
\date{\today}

\author{Miaomiao Yu}
\affiliation{College of Computer Science and Technology, %
National University of Defense Technology, %
Changsha 410073, China}%
\author{Jingyi Lv}
\affiliation{College of Advanced Interdisciplinary Studies, %
National University of Defense Technology, %
Changsha 410073, China}%
\author{Yan Wang}
\affiliation{College of Computer Science and Technology, %
National University of Defense Technology, %
Changsha 410073, China}%
\author{Kun Wang}
\email{nju.wangkun@gmail.com}
\affiliation{College of Computer Science and Technology, %
National University of Defense Technology, %
Changsha 410073, China}%
\author{Ping Xu}
\email{pingxu520@nju.edu.cn}
\affiliation{College of Computer Science and Technology, %
National University of Defense Technology, %
Changsha 410073, China}%

\begin{abstract}
Bipartite Gaussian boson sampling (BipartiteGBS) produces output probabilities governed by squared permanents of submatrices of arbitrary complex matrices, matching the nonsymmetric structure of directed graphs. Most GBS-based graph algorithms, however, rely on symmetric hafnian structure and are formulated for undirected problems. Here we propose a BipartiteGBS-based framework for directed-graph heuristic optimization. We introduce Max-Perm as a canonical optimization task for BipartiteGBS and derive a closed-form sampling enhancement factor relative to uniform classical sampling in this idealized setting. 
We then use permanent-biased BipartiteGBS samples to guide a genetic algorithm for 
the celebrated directed Hamiltonian cycle problem. 
Numerical experiments on Erd\H{o}s--R\'enyi random directed graphs show that the resulting BipartiteGBS-enhanced algorithms improve success rates over a standard genetic algorithm and yield longer valid paths when no Hamiltonian cycle is found, while ablation tests indicate that BipartiteGBS-guided initialization is the dominant contributor. These results show how permanent-based photonic sampling can provide useful algorithmic guidance for asymmetric combinatorial search.
\end{abstract}
\maketitle

\section{Introduction}\label{sec:introduction}

\begin{figure*}[t]
\centering
\includegraphics[width=0.9\linewidth]{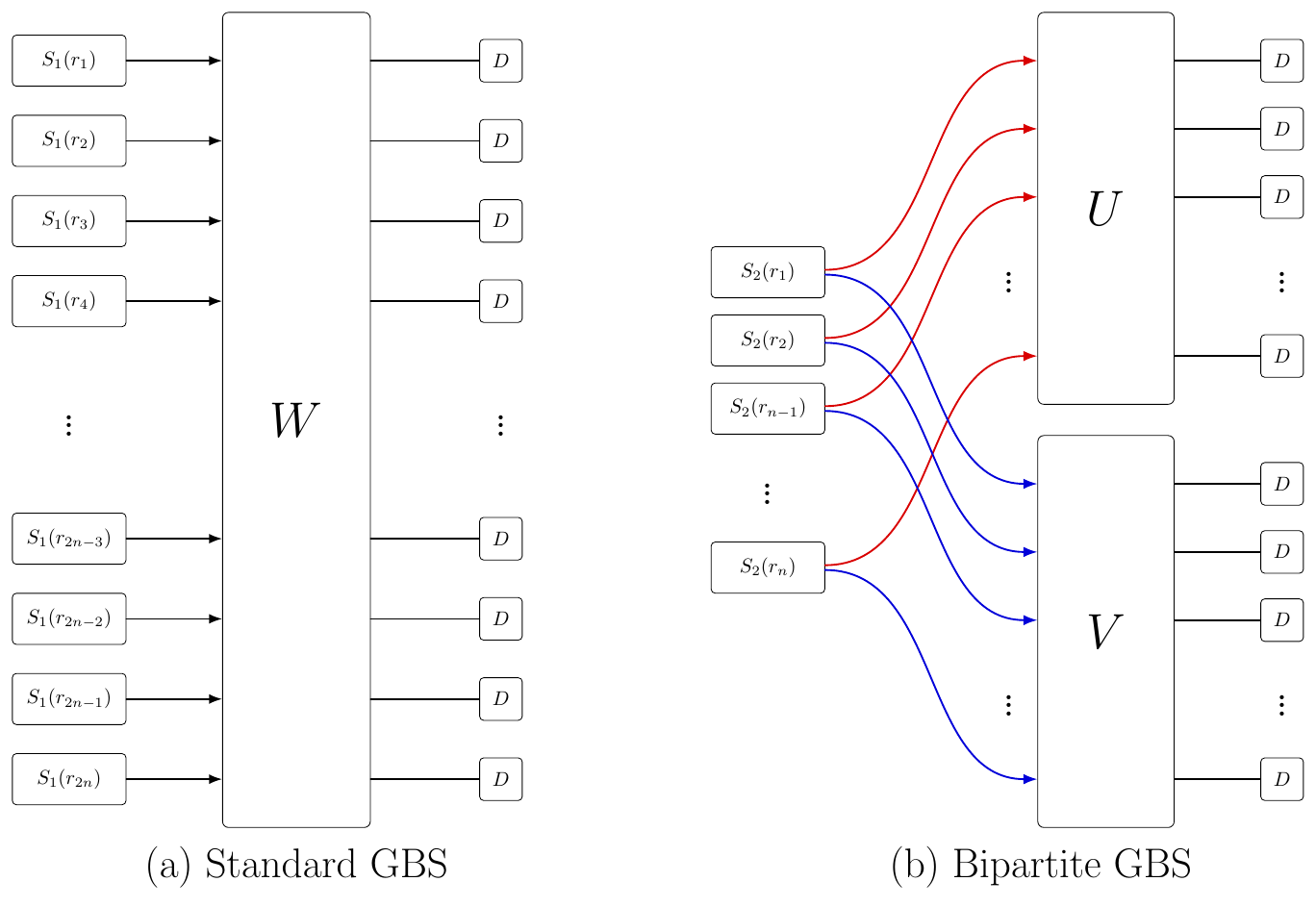}
\caption{\textbf{Comparison of standard GBS and BipartiteGBS optical architectures.}
(a) Standard GBS injects $2n$ single-mode squeezed states $S_1(r_i)$
into a single interferometer $W$, followed by photon-number detection.
(b) BipartiteGBS injects $n$ two-mode squeezed states $S_2(r_i)$;
the two output arms are sent through separate interferometers $U$
and $V$ and measured independently. Under the SVD convention used in
Sec.~\ref{sec:gbs}, the lower interferometer implements $V^*$; for the
real matrices considered here, $V^*=V$.
In the BipartiteGBS setting, the corresponding sampling probabilities are
proportional to squared permanents of submatrices of the matrix encoded by
the two interferometers and the squeezing parameters.}
\label{fig:bipartite-gbs-illustration}
\end{figure*}

Boson sampling and its Gaussian variants have become prominent platforms for probing photonic quantum advantage, because they connect experimentally accessible optical transformations with matrix functions whose exact sampling is believed to be classically hard. In the original boson sampling model~\cite{aaronson2011computational}, this hardness is associated with permanents of submatrices of a linear-optical unitary, and the model has motivated a series of experimental demonstrations~\cite{spring2013boson, broome2013photonic, crespi2013integrated, tillmann2013experimental} and related sampling variants~\cite{lund2014boson, barkhofen2017driven}. Gaussian boson sampling (GBS)~\cite{hamilton2017gaussian} addresses part of the experimental difficulty of single-photon boson sampling by replacing single photons with Gaussian states, which can be prepared deterministically and implemented on programmable photonic platforms~\cite{kruse2019detailed,zhong2020quantum,madsen2022quantum,liu2026gaussian}; its output probabilities are governed by Hafnians, leading to complexity-theoretic conjectures tailored to GBS~\cite{hamilton2017gaussian, kruse2019detailed}. Bipartite Gaussian boson sampling (BipartiteGBS)~\cite{grier2022complexity,arrazola2021quantum,borghi2025bipartite} further adapts this program to two-mode-squeezing platforms such as spontaneous parametric down-conversion and spontaneous four-wave mixing~\cite{arrazola2021quantum,wu2022optimization}, while retaining the same sampling-hardness conjecture as boson sampling~\cite{grier2022complexity}. As illustrated in Fig.~\ref{fig:bipartite-gbs-illustration}, standard GBS sends single-mode squeezed states through one interferometer, whereas BipartiteGBS uses two-mode squeezed states and two interferometers, making explicit the architectural distinction used throughout this work.

The matrix function and its symmetry structure become central when GBS is used for graph-theoretic applications. Existing GBS-based algorithms have been developed for graph problems~\cite{arrazola2018using, schuld2020measuring, bradler2018gaussian,zhu2025solving}, as well as related applications such as point processes~\cite{jahangiri2020point} and molecular vibronic spectra~\cite{huh2017vibronic}. However, standard GBS naturally encodes symmetric matrices and has therefore been used mainly for undirected graph problems. Directed graphs are ubiquitous in real-world systems, including social networks~\cite{wasserman1994social}, biological regulatory circuits~\cite{alon2019introduction}, and transportation infrastructures~\cite{bang2000theory}, precisely because they naturally encode asymmetric relationships. In such frameworks, the existence of a feasible or preferential transition from one vertex to another does not necessitate the reverse, a property that is fundamental to capturing directed influence, causality, and flow. This asymmetry renders directed graphs indispensable for modeling processes where directionality carries intrinsic semantic or physical meaning. 
Although directed graphs can in principle be embedded into larger symmetric matrices, such embeddings obscure the native nonsymmetric structure of a directed adjacency matrix. 
BipartiteGBS output probabilities are proportional to squared permanents of submatrices of arbitrary matrices, rather than being restricted to symmetric matrices. This makes BipartiteGBS a natural candidate for directed graphs; however, to the best of our knowledge, no BipartiteGBS-based algorithm has yet been proposed specifically for directed graph problems.

We choose the directed Hamiltonian cycle problem as a concrete testbed because it is both structurally asymmetric and computationally demanding. It is one of Karp's original 21 NP-complete problems~\cite{karp2009reducibility}: given a directed graph, the task is to determine whether there exists a cycle that visits every vertex exactly once and returns to the starting point~\cite{gould1991updating}. It is also closely related to the traveling salesman problem~\cite{matai2010traveling}. Since no polynomial-time algorithm is known for general directed graphs, heuristic and stochastic methods, including quantum-inspired approaches~\cite{lucas2014ising}, remain important for exploring practical performance.

\textbf{Contributions.} 
We propose a BipartiteGBS-enhanced genetic algorithm for directed Hamiltonian cycle search. 
The central design principle is to use BipartiteGBS as a source of permanent-biased heuristic information: sampled vertex substructures and edge co-occurrence frequencies are converted into guidance for a classical genetic algorithm. 
Numerical results on Erd\H{o}s--R\'{e}nyi directed graphs show improved search performance compared with a standard genetic algorithm, and ablation tests identify BipartiteGBS-guided initialization as the most effective component. 
These results indicate that heuristic information extracted from BipartiteGBS can enhance classical search for asymmetric combinatorial problems. 
This enhancement is heuristic in nature and does not constitute a claim of complexity-theoretic quantum advantage.

\textbf{Organization.} 
Sec.~\ref{sec:gbs} reviews the theoretical formalism of GBS and BipartiteGBS. 
Sec.~\ref{sec:approximate optimization} analyzes the enhancement ratio of BipartiteGBS for the Max-Perm problem. 
Sec.~\ref{sec:hc} describes the BipartiteGBS-enhanced genetic algorithm (\BipartiteGBSGA{}) for Hamiltonian cycle search,
including \MSBipartiteGBSGA{}.
Sec.~\ref{sec:numerical-simulations} presents numerical simulations and discussions.

\section{GBS and BipartiteGBS}
\label{sec:gbs}

In a standard GBS setup~\cite{hamilton2017gaussian, kruse2019detailed}, $2m$ single-mode squeezed vacuum states with squeezing parameters $\{r_1, \ldots, r_{2m}\}$ are injected into a linear interferometer $T$, as illustrated in Fig.~\ref{fig:bipartite-gbs-illustration}(a). All $2m$ output modes are measured by photon-number-resolving detectors. The probability of obtaining a photon-number pattern $\bar{c} = (c_1, \ldots, c_{2m})$ is given by
\begin{align}
\Pr(\bar{c}) = \frac{1}{\bar{c}! \sqrt{|\sigma_Q|}}\operatorname{Haf}(A_S).
\label{eq:pr_haf}
\end{align}
Here $A$ is the adjacency matrix of the Gaussian state, which takes the block-diagonal form $A = B \oplus B^*$,
$\sigma_Q$ is the covariance matrix of the Gaussian state,
and $\operatorname{Haf}$ is the Hafnian function. The $m \times m$ complex symmetric matrix $B$ encodes the interferometer $W$ and squeezing parameters $\{r_i\}$ through the Takagi decomposition~\cite{hamilton2017gaussian}
\begin{align}
    B = W\left(\operatorname{diag}(\tanh r_1, \ldots, \tanh r_{m})\right)W^T,
    \label{eq:takagi}
\end{align}
where $\tanh r_i \in [0, 1)$ for all $i$. 
Using the identity $\operatorname{Haf}(B \oplus B^*) = |\operatorname{Haf}(B)|^2$, Eq.~\eqref{eq:pr_haf} simplifies to
\begin{align}
    \Pr(\bar{c})=\frac{1}{\bar{c}!\sqrt{|\sigma_Q|}}|\operatorname{Haf}(B_S)|^2.
    \label{eq:haf_simple}
\end{align}
Here, $B_S$ denotes the submatrix of $B$ formed by selecting the rows and columns corresponding to the occupied output modes, and $\bar{c}! = \prod_{j=1}^{2m} c_j!$ accounts for photon-number multiplicities. 
When all $c_j \in \{0,1\}$, we have $\bar{c}! = 1$; this is the collision-free condition.

As illustrated in Fig.~\ref{fig:bipartite-gbs-illustration}(b), BipartiteGBS, introduced in Ref.~\cite{grier2022complexity}, is a specific GBS programming strategy that yields output probabilities proportional to the permanents of submatrices of \textit{arbitrary} complex matrices, after a possible global rescaling that makes the encoded matrix physically admissible. This is made possible by the identity
\begin{align}
    \operatorname{Per}(C)=\operatorname{Haf}\left[\begin{pmatrix}
        0 & C\\
        C^T & 0
    \end{pmatrix}\right], 
    \label{eq:haf_per_identity}
\end{align}
where $\operatorname{Per}$ is the Permanent function. This identity expresses the permanent of any matrix $C$ as the Hafnian of a 
block-antidiagonal symmetric matrix.

By specializing $B$ to the block-antidiagonal form
\begin{align}
    B=\begin{pmatrix}
    0 & C\\
    C^T & 0
\end{pmatrix}, 
    \label{eq:B_antidiag}
\end{align}
the Hafnian in Eq.~\eqref{eq:haf_simple} is converted into a permanent via Eq.~\eqref{eq:haf_per_identity}.
This block-antidiagonal $B$ is realized by programming the BipartiteGBS device as follows. Perform the singular value decomposition $C = U \Sigma V^{\dagger}$, where $\Sigma = \operatorname{diag}(\sigma_1,\ldots,\sigma_n)$ with $\sigma_i = \tanh r_i \in [0,1)$ for all $i$. Thus the encoded matrix must satisfy $\|C\|_2<1$; if the target matrix does not, it is first multiplied by a global scaling factor, which leaves relative probabilities among equal-size submatrices unchanged. In the convention of Fig.~\ref{fig:bipartite-gbs-illustration}(b), the upper interferometer is set to $U$, while the lower interferometer labeled $V$ implements the physical unitary $V^*$. The $n$ two-mode squeezing parameters are set to $\{r_i\}$. The singular values $\sigma_i = \tanh r_i$ are thus directly controlled by the squeezing strengths, while $U$ and $V^\dagger$ shape which submatrices of $C$ receive large permanent weight. For the real adjacency matrices used in this work, $U$ and $V$ are real orthogonal, hence $V^* = V$.
This construction generalizes scattershot boson sampling~\cite{lund2014boson} and its twofold variant~\cite{chakhmakhchyan2017boson}. 
Concretely, let $\mathbf{s}=(s_1,\ldots,s_n)$ and $\mathbf{t}=(t_1,\ldots,t_n)$ denote 
the photon-number patterns in the two output registers. The output probability becomes
\begin{align}
    \Pr(\mathbf{s},\mathbf{t})=\frac{1}{\sqrt{|\sigma_Q|}\prod_i s_i!\prod_j t_j!}\left|\operatorname{Per}(C_{\mathbf{s},\mathbf{t}})\right|^2, 
    \label{eq:pr_permanent}
\end{align}
where $C_{\mathbf{s},\mathbf{t}}$ is obtained from $C$ by repeating row $i$ exactly $s_i$ times and column $j$ exactly $t_j$ times. In the collision-free balanced case, $\mathbf{s}$ and $\mathbf{t}$ reduce to occupied mode sets $S$ and $T$ of equal size, the multiplicity factor satisfies $\prod_i s_i!\prod_j t_j!=1$, and $C_{\mathbf{s},\mathbf{t}}=C_{S,T}$.

\section{Quantum Approximate Optimization with BipartiteGBS}\label{sec:approximate optimization}

A key property of BipartiteGBS is that its output probabilities are proportional to the permanents of submatrices of arbitrary matrices. Leveraging this property, we analyze an idealized sampling enhancement and then use it heuristically. This generalizes the quantum approximate optimization framework based on standard GBS~\cite{arrazola2018quantum} to the BipartiteGBS regime. 
Analogous to the Max-Haf problem central to that framework, 
we introduce the following Max-Perm problem.

\begin{figure}[t]
\centering
\includegraphics[width=0.9\linewidth]{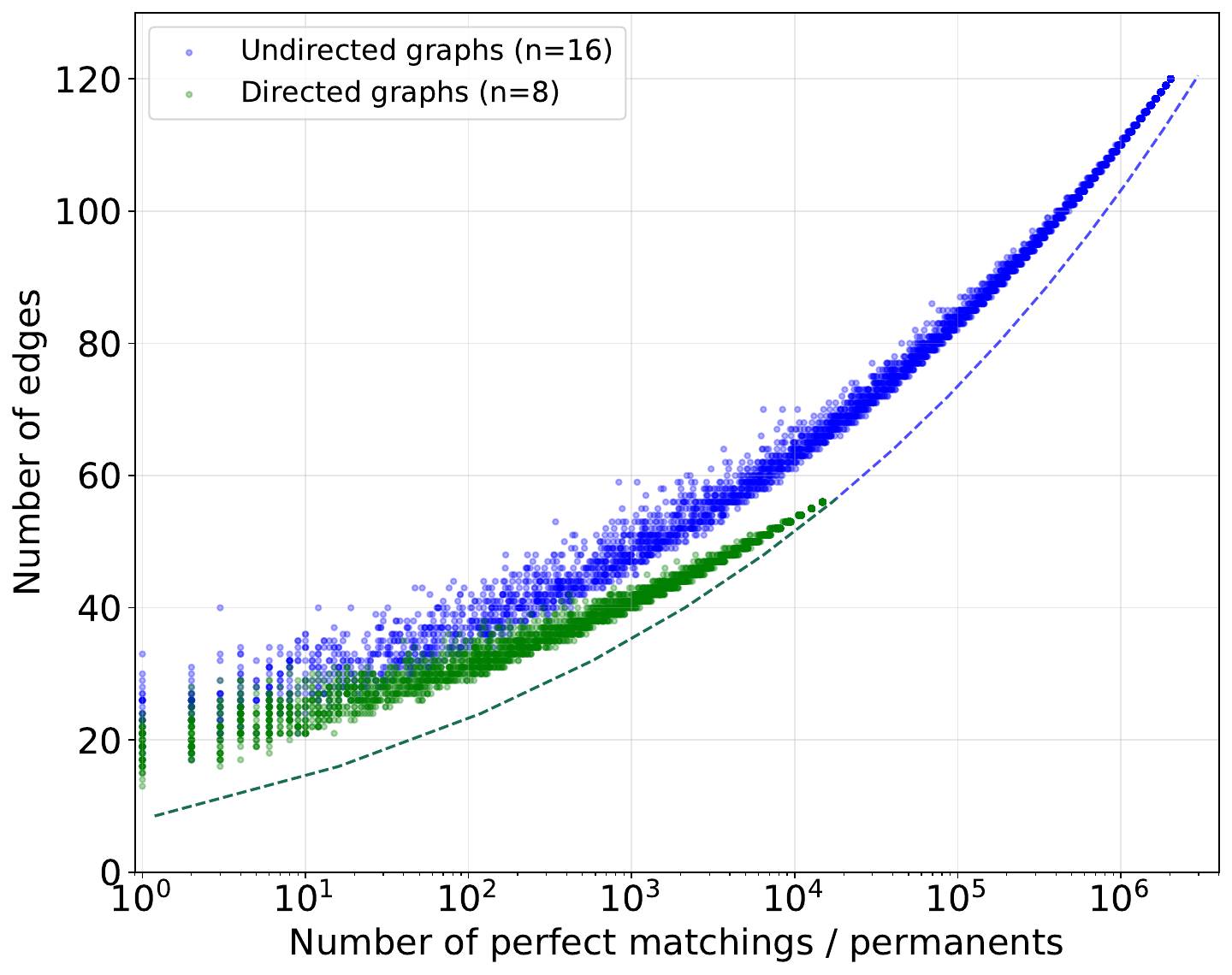}
\caption{\textbf{Scatter plots relating edge count to matching-related matrix functions.}
Blue dots (left panel) represent random undirected graphs on $16$ vertices; edge count is plotted against the number of perfect matchings, equivalently the Hafnian of the adjacency matrix. Green dots (right panel) represent random directed graphs on $8$ vertices; edge count is plotted against the permanent of the adjacency matrix. Blue and green dashed curves indicate the theoretical upper bounds from Eq.~\eqref{upper bound} and Eq.~\eqref{upper bound for per}, respectively.
For each graph type, 300 random graphs are generated at each of the edge-probability values
$p = 0.20, 0.25, \ldots, 0.95, 0.97, 0.98, 0.99, 0.995, 0.999$ (21 values).}
\label{fig:relation}
\end{figure}

\begin{definition}[Max-Perm problem]
Given a matrix $C \in \mathbb{C}^{n \times n}$ and a fixed integer $k$, the Max-Perm problem seeks the submatrix $C_{S,T}$ that maximizes $|\operatorname{Per}(C_{S,T})|^2$:
\begin{align}
    \arg\max_{C_{S,T}} \left|\operatorname{Per}(C_{S,T})\right|^2,
\end{align}
subject to $\sum_i s_i = \sum_j t_j = k$.
\end{definition}

The NP-hardness of Max-Perm follows from a reduction of the Balanced Biclique problem. We consider a worst-case setting. Given a bipartite graph $G$ with $n$ vertices on each side, construct its $n \times n$ biadjacency matrix $C$. For any $k$, a $k \times k$ submatrix $C_{S,T}$ is the all-ones matrix if and only if $G[S,T]$ is a complete bipartite graph $K_{k,k}$, in which case $\operatorname{Per}(C_{S,T}) = k!$. Any submatrix missing at least one edge has permanent strictly less than $k!$. So decide whether the optimum over $k\times k$ submatrices reaches $(k!)^2$ (or $k!$ before squaring), which is equivalent to the existence of a $K_{k,k}$. As the decision problem of Balanced Biclique is NP-complete~\cite{garey1979computers}, Max-Perm is NP-hard as well.

A straightforward classical random search for the Max-Perm problem proceeds by uniformly sampling $k$-dimensional submatrices $C_{S,T}$ of $C$. For each sample, the permanent is evaluated, and the submatrix yielding the largest value is retained. The set of all $k$-dimensional submatrices of $C$ is labeled by
\begin{align}
    \Gamma_{S,T,k} = \{(S,T): s_i, t_j \in \{0,1\},\; \textstyle\sum_i s_i = \sum_j t_j = k\},
\end{align}
where $(S,T) \in \Gamma_{S,T,k}$ specifies the selected rows and columns. Since $|\Gamma_{S,T,k}| = \binom{n}{k}^2$, the expectation value of $|\operatorname{Per}(C_{S,T})|^2$ under uniform sampling is
\begin{align}
    \hat{\mu}_U = \binom{n}{k}^{-2} \sum_{(S,T)\in \Gamma_{S,T,k}} |\operatorname{Per}(C_{S,T})|^2.
\end{align}

Alternatively, drawing $C$ from the distribution $\mathcal{G}: \mathcal{N}\bigl(0, \frac{\tanh^2 r}{n}\bigr)_{\mathbb{C}}^{n \times n}$~\cite{grier2022complexity} allows us to approximate this quantity as $\hat{\mu}_U \approx \mathbb{E}_{\mathcal{G}}[|\operatorname{Per}(C_{S,T})|^2]$.

We now compute the expectation of $|\operatorname{Per}(C_{S,T})|^2$ when employing BipartiteGBS, expressing the result in terms of $n$, $k$, and the squeezing parameter $r$. Let $p_{cf}$ denote the probability that a BipartiteGBS sample yields a collision-free configuration, i.e., a pair $(S,T) \in \Gamma_{S,T,k}$ with no two photons occupying the same mode. The collision-free probability approaches unity in the dilute regime $k^2/n \to 0$, rather than merely from $k\ll n$; under a uniform-mode approximation for the two BipartiteGBS registers, $p_{cf}\approx [\prod_{j=0}^{k-1}(1-j/n)]^2$. We therefore keep $p_{cf}$ explicit below. Furthermore, let $q_{n,r}(k)$ be the probability of generating exactly $k$ photon pairs from $n$ two-mode squeezed vacua with equal squeezing $r$. For a single two-mode squeezed vacuum, the photon-pair number follows a geometric distribution $P(k)=(1-\tanh^2r)(\tanh^2r)^k$~\cite{perrier2019thermal}. For $n$ independent sources, the total number of photon pairs follows the negative binomial distribution
\begin{align}
    q_{n,r}(k) = \binom{n+k-1}{k} (\operatorname{sech}^2 r)^n (\tanh^2 r)^k.
    \label{q probability}
\end{align}

For all collision-free configurations $(S,T) \in \Gamma_{S,T,k}$, we have $\prod_i s_i! \prod_j t_j! = 1$, which simplifies the relation between sampling probability and permanent. Consequently, the expectation of $|\operatorname{Per}(C_{S,T})|^2$ under BipartiteGBS satisfies
\begin{align}
        \hat{\mu}_{\text{BipartiteGBS}} &= \frac{1}{p_{cf} \cdot q_{n,r}(k) \cdot (\cosh^2 r)^n} \sum_{S,T} |\operatorname{Per}(C_{S,T})|^4 \notag\\
        &\approx \frac{\binom{n}{k}^2}{p_{cf} \cdot q_{n,r}(k) \cdot (\cosh^2 r)^n} \, \mathbb{E}_{\mathcal{G}}\bigl[|\operatorname{Per}(C_{S,T})|^4\bigr],
\end{align}
where the summation is over all $(S,T) \in \Gamma_{S,T,k}$. In the comparison below we condition on generating exactly $k$ photon pairs and set $p_{cf}=1$ only for an idealized, postselected comparison with Ref.~\cite{arrazola2018quantum}; raw-shot performance would include the fixed-$k$ and collision-free postselection overhead $1/[q_{n,r}(k)p_{cf}]$.

To evaluate the expectation, we recall known results for the moments of permanents of Gaussian random matrices. Define the rescaled matrix
\begin{align}
    X = \frac{\sqrt{n}}{\tanh r} C_{S,T},
\end{align}
so that $X \sim \mathcal{N}(0,1)_{\mathbb{C}}^{k \times k}$ is a standard complex Gaussian matrix. For such a matrix, Ref.~\cite{aaronson2011computational} provides the following moments of the permanent:
\begin{align}
        &\mathbb{E}_{\mathcal{G}}\bigl[|\operatorname{Per}(X)|^2\bigr] = k!, \notag\\
        &\mathbb{E}_{\mathcal{G}}\bigl[|\operatorname{Per}(X)|^4\bigr] = (k+1)(k!)^2.
\end{align}

Transforming back to $C_{S,T}$ yields
\begin{align}
        &\mathbb{E}_{\mathcal{G}}\bigl[|\operatorname{Per}(C_{S,T})|^2\bigr] = \frac{(\tanh^2 r)^k \, k!}{n^k}, \notag\\
        &\mathbb{E}_{\mathcal{G}}\bigl[|\operatorname{Per}(C_{S,T})|^4\bigr] = \frac{(\tanh^4 r)^k \, (k+1)(k!)^2}{n^{2k}}.
\end{align}

Substituting the fourth-moment result into the expression for $\hat{\mu}_{\text{BipartiteGBS}}$, conditioning on exactly $k$ photon pairs, and taking the collision-free postselected convention $p_{cf}=1$, we obtain the enhancement factor for our BipartiteGBS-based Max-Perm solver:
\begin{align}
    R_{\operatorname{Per}} = \frac{\hat{\mu}_{\text{BipartiteGBS}}}{\hat{\mu}_U} = \frac{\binom{n}{k}^2 (k+1)k!}{\binom{n+k-1}{k} n^k}.
\end{align}

This factor quantifies the fixed-$k$, collision-free postselected sampling enhancement of BipartiteGBS over uniform classical sampling for the Max-Perm problem. For comparison, Tab.~\ref{table1} presents the enhancement factor $R_{\operatorname{Per}}$ alongside the Hafnian-based enhancement $R_{\operatorname{haf}}$ reported in Ref.~\cite{arrazola2018quantum} for the case $n = k^2$. The table should therefore be read as an idealized comparison of sampling bias; an unpostselected implementation must also account for the fixed-$k$ and collision-free probability factor $q_{n,r}(k)p_{cf}$. Our results demonstrate that BipartiteGBS provides a significant and growing sampling enhancement for permanent-based optimization as $k$ increases. 

This framework exemplifies proportional sampling for optimization and demonstrates how BipartiteGBS can enhance classical stochastic algorithms. The Max-Perm problem serves as the canonical optimization task for BipartiteGBS, providing the essential insights for understanding its application to approximate optimization. We now unlock the broader potential of BipartiteGBS by carrying these insights over to a wider class of optimization problems, specifically, the directed Hamiltonian cycle problem.

\begin{table}[htbp]
\centering
\setlength{\tabcolsep}{0pt} 
\setlength\heavyrulewidth{0.3ex}  
\renewcommand{\arraystretch}{1.5} 
\begin{tabular}{@{}w{c}{0.15\textwidth}w{c}{0.15\textwidth}w{c}{0.15\textwidth}@{}}
\toprule
$k$ & $R_{\text{haf}}$ & $R_{\text{per}}$ (postselected) \\
\midrule
$4$   & $1.58$ & $1.56$ \\
$8$   & $1.93$ & $2.37$ \\
$10$  & $2.10$ & $2.80$ \\
$20$  & $2.81$ & $5.00$ \\
$40$  & $3.85$ & $9.46$ \\
$100$ & $5.99$ & $22.84$ \\
\bottomrule
\end{tabular}
\caption{\textbf{Comparison of Hafnian- and permanent-based enhancement factors.}
The table lists $R_{\operatorname{haf}}$ from Ref.~\cite{arrazola2018quantum} and $R_{\operatorname{per}}$ from this work for $n = k^2$, using the fixed-$k$, collision-free postselected convention.}
\label{table1}
\end{table}

\section{BipartiteGBS for Searching Hamiltonian Cycles}
\label{sec:hc}

In this section, we move from the idealized Max-Perm problem studied in Sec.~\ref{sec:approximate optimization} to a concrete directed-graph search problem: the Hamiltonian cycle problem. The connection is through the permanent-biased sampling mechanism of BipartiteGBS. Sec.~\ref{sec:approximate optimization} shows that BipartiteGBS preferentially samples submatrices with large squared permanent and quantifies this bias for Max-Perm. When the encoded matrix is the adjacency matrix of a directed graph, these permanents have a direct graph-theoretic interpretation: the permanent of a principal submatrix counts directed cycle covers on the selected vertices, equivalently perfect matchings in the corresponding bipartite transformation.

Hamiltonian cycle search is stricter than Max-Perm. A Hamiltonian cycle is a single directed cycle visiting all vertices exactly once, whereas a large permanent may arise from many disjoint cycle covers or dense local structures. Therefore, Sec.~\ref{sec:approximate optimization} does not provide a direct solver or complexity-theoretic advantage for Hamiltonian-cycle search. Instead, it supplies the sampling principle used here: high-permanent substructures are treated as informative local evidence about promising vertices and directed edges. We convert BipartiteGBS samples into sampled subgraphs and empirical edge frequencies, and use this heuristic information to guide a classical genetic algorithm. In the experimental setting, sampled row and column registers are further abstracted into a single vertex set $S\cup T$. This abstraction retains directed-edge candidates inside the union but is not distributed as a principal-submatrix sample. We emphasize that the projection is a heuristic post-processing step. This heuristic is empirically justified by its performance in our experiments, but does not carry theoretical guarantees. For applications requiring a theoretically grounded distribution over single vertex sets, one can restrict to cases where $S=T$, in which case the projected set $S$ inherits the probability proportional to $|\operatorname{Per}(A_{S,S})|^2$.

The remainder of this section makes this heuristic pipeline explicit. We first describe how BipartiteGBS encodes a matrix and produces postselected samples, and then introduce the permanent-guided heuristic, the genetic algorithm, 
and the multi-stage variant.


\subsection{Directed graph encoding and sampling}\label{sec:encoding-and-sampling}

We now describe how a real $n \times n$ matrix $A$, typically the adjacency matrix of a directed graph, is encoded into a BipartiteGBS device and how the resulting samples are postselected and converted into heuristic information. The full pipeline is specified in Alg.~\ref{alg:gbs_sampler_simple}.

\textbf{Encoding.}
Recall from Sec.~\ref{sec:gbs} that BipartiteGBS realizes $C = U \Sigma V^\dagger$ with $\Sigma = \operatorname{diag}(\sigma_1, \ldots, \sigma_n)$, $\sigma_i = \tanh r_i \in [0, 1)$. The encoding proceeds as follows:
\begin{enumerate}
    \item Compute the SVD $A = U \Sigma_{\text{raw}} V^\dagger$, where $U, V$ are real orthogonal and $\Sigma_{\text{raw}} = \operatorname{diag}(\sigma_1^{\text{raw}}, \ldots, \sigma_n^{\text{raw}})$.
    \item Rescale the singular values to lie in $[0, 1)$:
    \begin{align}
        \tilde{\sigma}_i = \eta \cdot \frac{\sigma_i^{\text{raw}}}{\max_j \sigma_j^{\text{raw}}}, \qquad \eta \in (0, 1).
        \label{eq:rescaling}
    \end{align}
    \item Set the squeezing parameters as $r_i = \operatorname{arctanh}(\tilde{\sigma}_i)$.
    \item Use $U$ and $V^* = V$ as the two interferometers.
\end{enumerate}
The physically encoded matrix is therefore $C = \frac{\eta}{\max_j \sigma_j^{\text{raw}}} A$, i.e., $A$ scaled by a global constant. For any $m \times m$ submatrix $C_{S,T}$, we have $|\operatorname{Per}(C_{S,T})|^2 = \alpha^{2m} |\operatorname{Per}(A_{S,T})|^2$ with $\alpha = \eta / \max_j \sigma_j^{\text{raw}}$. Since $\alpha^{2m}$ is constant for fixed $m$, the relative sampling bias among equal-size submatrices is preserved. The hyperparameter $\eta$ controls the photon-pair production rate $\mathbb{E}[N_{\text{pair}}] = \sum_i \sinh^2 r_i$; we use $\eta = 0.75$.

\textbf{Sampling and postselection.}
Each BipartiteGBS shot yields occupied mode sets $S$ (first register) and $T$ (second register) with $\Pr(S,T) \propto |\operatorname{Per}(C_{S,T})|^2$. We retain only \textit{informative samples} that are (i)~collision-free (no mode receives $>1$ photon, so $\prod_i s_i! \prod_j t_j! = 1$) and (ii)~balanced ($|S| = |T| \ge 3$).

\textbf{Heuristic post-processing.}
Each accepted sample is projected to a vertex pool $V_S = S \cup T$ (see Sec.~\ref{sec:hc} for the heuristic justification). Over $N_{\text{BipartiteGBS}}$ shots, the accepted pools yield (i)~a collection $\mathcal{S}$ of vertex subsets used to seed the genetic algorithm's initial population, and (ii)~empirical co-occurrence frequencies
\begin{align}
    p_{ed}(u,v) = \frac{\bigl|\{V_S \in \mathcal{S} : u, v \in V_S,\; (u,v) \in E\}\bigr|}{\max(|\mathcal{S}|,\, 1)},
    \label{eq:cooccurrence}
\end{align}
which serve as edge-importance weights in fitness evaluation and mutation.


\subsection{Permanent-guided heuristic}

\begin{figure*}[t]
\centering
\includegraphics[width=1.0\linewidth]{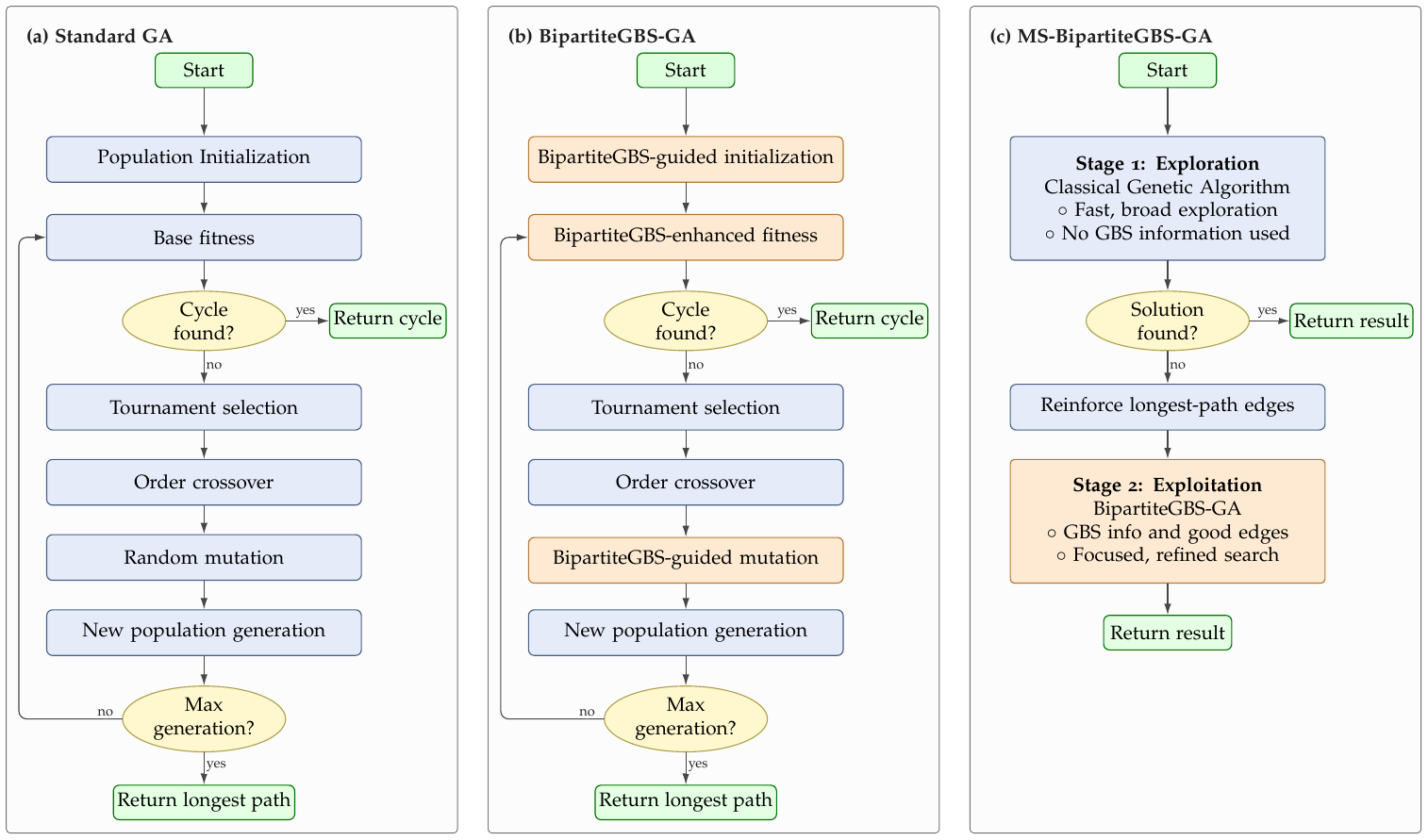}
\caption{\textbf{Workflows for classical and BipartiteGBS-guided Hamiltonian-cycle search.} 
(a) Standard genetic algorithm (GA);
(b) BipartiteGBS-enhanced genetic algorithm (\BipartiteGBSGA{});
(c) Multi-stage \BipartiteGBSGA{} (\MSBipartiteGBSGA{}).
Blue blocks indicate classical computational steps, whereas orange blocks indicate quantum sampling steps based on BipartiteGBS.}
\label{fig:flowcharts}
\end{figure*}

Classical algorithms for Hamiltonian cycle search, such as backtracking~\cite{reingold1977combinatorial, cormen2022introduction} and the Held-Karp dynamic programming approach~\cite{held1962dynamic}, share a common limitation: they scale exponentially with the number of vertices. Backtracking implicitly enumerates permutations, while Held-Karp enumerates subsets of vertices and endpoints, leading to a time complexity of $O(n^2 2^n)$. This exponential scaling motivates the exploration of stochastic and quantum-inspired approaches~\cite{corrente2023comparison}. 

To provide theoretical background for our heuristic, we first establish a connection between the permanent of a directed graph and its edge density. Intuitively, a graph with many perfect matchings is expected to contain many edges. This intuition was made quantitative in Ref.~\cite{aaghabali2015upper}, which also establishes a correspondence between the permanent of a directed graph and its edge density. Let $G=(V,E)$ be a simple undirected graph with an even number of vertices $2h := |V|$ and $e := |E|$ edges. An upper bound on the number of perfect matchings in $G$ is given by
\begin{align}
    \operatorname{PM}(G) \leq \left( \left\lfloor \frac{e}{h} \right\rfloor ! \right)^{\frac{h-a}{\left\lfloor \frac{e}{h} \right\rfloor}} \left( \left\lceil \frac{e}{h} \right\rceil ! \right)^{\frac{a}{\left\lceil \frac{e}{h} \right\rceil}},
    \label{upper bound}
\end{align}
where $a := e - h \left\lfloor \frac{e}{h} \right\rfloor$ and $\operatorname{PM}$ denotes the number 
of perfect matchings. If $e<h, \operatorname{PM}(G)=0$.

Consider a directed graph $G_D$ with $n$ vertices and $m$ edges, and let $C$ be its adjacency matrix. Ref.~\cite{aaghabali2015upper} constructs its bipartite transformation $B(G_D)$ by splitting each vertex $v_i$ into two copies $v_i'$ and $v_i''$, and adding an undirected edge $(v_i', v_j'')$ for every directed edge $(v_i, v_j)$ in $G_D$. The number of perfect matchings in $B(G_D)$ is
\begin{align}
    \operatorname{PM}(B(G_D)) = \operatorname{Per}(C).
\end{align}
Since $B(G_D)$ is a bipartite graph with $2n$ vertices and $m$ edges, applying the bound in Eq.~\eqref{upper bound} with the substitutions $e \to m$ and $h \to n$ yields
\begin{align}
    \operatorname{Per}(C)
\leq \left( \left\lfloor \frac{m}{n} \right\rfloor ! \right)^{\frac{n-a}{\left\lfloor \frac{m}{n} \right\rfloor}} \left( \left\lceil \frac{m}{n} \right\rceil ! \right)^{\frac{a}{\left\lceil \frac{m}{n} \right\rceil}},
    \label{upper bound for per}
\end{align}
where $a := m - n \left\lfloor \frac{m}{n} \right\rfloor$. If $m<n$, $\operatorname{Per}(C)=0$.
Thus, the permanent of an arbitrary directed graph obeys the same upper bound as the number of perfect matchings in an undirected graph with the same numbers of vertices and edges.

Fig.~\ref{fig:relation} illustrates the connection between matching-related matrix functions and edge density by showing the relationship between the number of perfect matchings and edge count for random undirected graphs on $16$ vertices, alongside the corresponding relationship between the permanent and edge count for random directed graphs on $8$ vertices. This motivates permanent-biased BipartiteGBS sampling as a heuristic for selecting dense directed substructures for the subsequent genetic search, without implying that large permanents certify Hamiltonicity.

Next, we recall a classical result linking graph density to Hamiltonicity. Since Dirac's 1952 theorem~\cite{dirac1952some}, a central line of research has been to identify sufficient conditions for Hamiltonicity, typically expressed in terms of edge density. A landmark result for directed graphs is the Ghouila-Houri theorem:

\begin{theorem}[Ghouila-Houri, 1960~\cite{ghouilahouri1960condition}]
\label{thm:dirac}
Every directed graph on 
$n$ vertices with minimum out-degree and in-degree at least $n/2$ contains a directed Hamiltonian cycle.
\end{theorem}

Under the Ghouila-Houri theorem, every vertex in a sufficiently dense directed graph has in-degree and out-degree at least $n/2$, guaranteeing the existence of a Hamiltonian cycle. In such dense graphs, one might expect that edges belonging to a Hamiltonian cycle are often embedded in locally dense substructures: for many cycle edges $(u,v)$, the large neighborhoods of $u$ and $v$ may overlap with those of adjacent vertices along the cycle, potentially creating locally dense subgraphs around those edges. While this observation does not constitute a rigorous proof and may not hold for every cycle edge, it provides a plausible heuristic basis for using locally dense substructures as reservoirs of cycle edges. We adopt this as a working hypothesis, leaving a rigorous theoretical characterization of the relationship between edge density and Hamiltonian cycle edges for future investigation.

BipartiteGBS provides two complementary types of information that align with this hypothesis. First, it generates accepted row/column samples that are projected to vertex pools. The underlying ordered pair $(S,T)$ is permanent-biased through $|\operatorname{Per}(A_{S,T})|^2$, whereas the projected pool $S\cup T$ is used heuristically as a candidate local structure rather than as a principal-submatrix sample. Second, from multiple projected samples we estimate co-occurrence frequencies: the empirical frequency with which both endpoints of a given directed edge appear together in a projected vertex pool while the edge itself exists in the original graph. Edges with higher co-occurrence frequencies tend to be embedded in locally dense substructures and, by heuristic extension, are more likely to belong to a Hamiltonian cycle.

\subsection{BipartiteGBS-enhanced genetic algorithm}

We next explain how this permanent-guided information is incorporated into a genetic algorithm.

The standard genetic algorithm~\cite{holland1992adaptation} encodes each possible vertex ordering as a chromosome and iteratively optimizes the population through selection, crossover, and mutation. The flowchart is given in Fig.~\ref{fig:flowcharts}(a). The BipartiteGBS-enhanced genetic algorithm (\BipartiteGBSGA{}),
as shown in Fig.~\ref{fig:flowcharts}(b), extends the classical framework by incorporating 
two forms of BipartiteGBS-derived information: 
\emph{co-occurrence frequencies}, which assign importance scores to directed edges; and 
\emph{sampled subgraphs}, which identify vertex subsets likely to contain useful cycle-cover structures. 
The proposed algorithm components are described below, with the full pseudocode provided 
in Alg.~\ref{alg:gbs_ga}.

\textbf{BipartiteGBS-guided initialization.} BipartiteGBS-guided initialization creates an initial population using three complementary strategies. A portion of individuals are generated as random permutations to maintain genetic diversity. 
Another portion are constructed from BipartiteGBS-sampled subgraphs, where each sample identifies a set of vertices that frequently co-appear in the quantum measurements, and a path is built by greedily selecting edges with the highest co-occurrence frequencies. 
The remaining individuals are created through a hybrid approach that combines BipartiteGBS top edges with random neighbor selection. The proportions are balanced to ensure that the population contains both diverse random solutions and a sufficient number of high-quality BipartiteGBS-informed solutions. This multi-strategy initialization enables the algorithm to benefit from quantum guidance while preserving exploratory capability. The pseudocode is provided in Alg.~\ref{alg:gbs_init}.

\textbf{BipartiteGBS-enhanced fitness.} The BipartiteGBS-enhanced fitness function evaluates an individual using two complementary criteria. The base fitness, denoted as $\text{Base}$, is the fraction of consecutive edges in the permutation that actually exist in the graph, i.e., $\text{Base} = \ell / n$, where $\ell$ is the number of valid edges and $n$ is the number of vertices. The BipartiteGBS contribution, denoted as $\text{BipartiteGBS}$, is the average co-occurrence frequency of these edges, reflecting their estimated importance according to BipartiteGBS sampling. This contribution is normalized to the interval $[0,1]$; its calculation is shown in the pseudocode in Alg.~\ref{alg:gbs_fitness}. These two criteria are combined as follows:
\begin{align}
    \text{Fitness} = \text{Base} \times (1 - \alpha) + \text{BipartiteGBS} \times \alpha,
\end{align}
where $\alpha$ controls the influence of BipartiteGBS information. To balance exploration and exploitation, $\alpha$ increases linearly from $0$ to $\alpha_{\max}$ over the course of evolution (in the main comparison of Fig.~\ref{fig:algorithm comparition}, $\alpha_{\max}=0.1$). In the sensitivity analysis that follows, we evaluate fixed $\alpha$ configurations to isolate the effect of this hyperparameter, where the fixed value corresponds to the value of $\alpha_{\max}$ used in the dynamic schedule. In early generations, $\alpha$ is small, making the algorithm rely primarily on base fitness for broad exploration. In later generations, $\alpha$ grows, giving more weight to BipartiteGBS-derived co-occurrence information to refine promising solutions.

\textbf{BipartiteGBS-guided mutation.} The BipartiteGBS-guided mutation operator applies mutation with a focus on repairing low-quality edges. With $70\%$ probability, it uses BipartiteGBS information to guide the mutation: first, it calculates the co-occurrence frequency for each edge in the circular path. It then identifies the edge with the lowest frequency, representing the most suspicious or unlikely connection. This edge is targeted by swapping one of its endpoint vertices with a randomly chosen vertex in the permutation, potentially replacing the low-probability edge with a better one. The remaining $30\%$ of the time, the operator falls back to traditional random swap mutation to maintain exploration diversity. This hybrid approach ensures that mutations are both guided by quantum information and sufficiently random to avoid premature convergence. The pseudocode is provided in Alg.~\ref{alg:gbs_mutation}.

\subsection{Multi-stage \BipartiteGBSGA{} (\MSBipartiteGBSGA{})}\label{sec:multi-stage-variant}

\MSBipartiteGBSGA{} (Alg.~\ref{alg:ms_bipartitegbs_ga}) adopts a two-stage serial
strategy as shown in Fig.~\ref{fig:flowcharts}(c). It uses a total generation budget $G_{\max}^{\mathrm{MS}}=300$, allocated as follows. Stage~1 performs a fast classical genetic algorithm with half the population ($N_{\mathrm{pop}}/2=50$) for $G_{\max}^{\mathrm{MS}}/3=100$ generations. If a Hamiltonian cycle is found, it returns immediately. If not, it extracts the longest path found and boosts the co-occurrence frequencies of edges along this path to $0.8$. This value is a heuristic reinforcement level below $1$, chosen to favor the best classical path without making those edges deterministic. Stage~2 then runs the full \BipartiteGBSGA{} with the full population ($N_{\mathrm{pop}}=100$) for $G_{\max}^{\mathrm{MS}}/2=150$ generations. The total evaluations across both stages equal $50\times100 + 100\times150 = 20\,000 = N_{\mathrm{pop}}G_{\max}$, matching the evaluation budget of the baseline GA ($N_{\mathrm{pop}}=100$, $G_{\max}=200$). This ``fast-first, precise-later'' design allows the algorithm to rapidly solve easy instances while deploying BipartiteGBS resources only when necessary.

\section{Numerical Simulations}\label{sec:numerical-simulations}

\subsection{Numerical Setup}\label{sec:numerical-setup}

We compare the following five algorithms, each specified by an algorithm block in the appendices:
\begin{itemize}
    \item the nearest neighbor heuristic (NN), described in Alg.~\ref{alg:nn_baseline} of Appx.~\ref{appx:classical-benchmarks};
    \item a standard genetic algorithm (GA), described in Alg.~\ref{alg:standard_ga} of Appx.~\ref{appx:classical-benchmarks};
    \item Degree-GA, the degree-centrality baseline briefly described below; its full procedure is given in Alg.~\ref{alg:degree_ga} of Appx.~\ref{appx:classical-benchmarks};
    \item \BipartiteGBSGA{},
            described in Alg.~\ref{alg:gbs_ga} of Appx.~\ref{appx:BipartiteGBS-GA}; and
    \item \MSBipartiteGBSGA{}, described in Alg.~\ref{alg:ms_bipartitegbs_ga} of Appx.~\ref{appx:MS-BipartiteGBS-GA}.
\end{itemize}

The NN heuristic implemented here constructs a path by starting from up to $10$ randomly chosen vertices
and repeatedly selecting the smallest indexed unvisited outgoing neighbor.
The GA follows a standard permutation based encoding with ordered crossover and random swap mutation. They are classical baselines.

\textbf{Degree-centrality baseline.}
To investigate whether the information provided by BipartiteGBS is qualitatively different from simple vertex-level heuristics, we compare \BipartiteGBSGA{} against a degree-centrality-enhanced genetic algorithm (Degree-GA). In Degree-GA, each directed edge $(u,v)$ is assigned a weight
\begin{align}
    w_{u,v}=\frac{1}{2}
    \left(\frac{\deg^+(u)}{\max_{x}\deg^+(x)}+\frac{\deg^-(v)}{\max_{x}\deg^-(x)}\right),
\end{align}
where $\deg^+(u)$ and $\deg^-(v)$ denote the out-degree of $u$ and the in-degree of $v$, respectively. This design is motivated by the structure of directed Hamiltonian cycles: entering a vertex $v$ requires sufficient in-degree, while leaving a vertex $u$ requires sufficient out-degree. The normalization ensures that weights lie in $[0,1]$ for consistent comparison with BipartiteGBS-derived co-occurrence frequencies, and the average balances both contributions. These weights are used to bias the initialization, fitness, and mutation operators, analogous to how the co-occurrence frequencies guide the search in \BipartiteGBSGA{}. Degree centrality captures only individual vertex statistics: it favors vertices with many connections but ignores how vertices co-occur within local substructures.

\textbf{Informative samples.}
In the directed graph encoding and sampling pipeline (Alg.~\ref{alg:gbs_sampler_simple}), an informative sample requires collision-free detection and balanced detection with $|S| = |T| \ge 3$. Samples not meeting both criteria are discarded. We use $N_{\text{BipartiteGBS}}=500$ shots per graph instance across all tested sizes ($n=15,20,25,30,35,40$).
For a graph with singular values $\{\sigma_i\}$, we encode the normalized singular values $\tilde{\sigma}_i=\eta\sigma_i/\max_j\sigma_j$ with $\eta=0.75$ and set the squeezing parameters as $r_i = \operatorname{arctanh}(\tilde{\sigma}_i)$. The expected total photon-pair number is $\mathbb{E}[N_{\text{pair}}] = \sum_{i=1}^n \sinh^2 r_i$. The accepted-sample fraction therefore depends on the full normalized singular-value spectrum of each graph instance, as well as on the collision-free and balanced-photon postselection conditions. Because the simulations did not log accepted-sample fractions for each instance, we do not infer a monotonic scaling law for this fraction. The verified fact used in the numerical study is that the $500$-shot budget produced at least one informative sample for every tested graph instance.

\textbf{Benchmark instances and simulation settings.}
To evaluate scalability, we test Erd\H{o}s--R\'enyi random graphs with edge probability $p=0.3$ and sizes $n=15,20,25,30,35,40$. BipartiteGBS samples are generated using Strawberry~Fields~\cite{killoran2019strawberry} with the Gaussian backend; the graph adjacency matrix $A$ is decomposed via SVD, the encoded singular values are set to $\tilde{\sigma}_i=\eta\Sigma_{ii}/\max_j\Sigma_{jj}$, and the squeezing parameters are set to $r_i = \operatorname{arctanh}(\tilde{\sigma}_i)$. For each graph instance, we draw $N_{\text{BipartiteGBS}}=500$ shots and retain informative samples according to the postselection rule defined above. For fair comparison, the GA-based algorithms (GA, Degree-GA, \BipartiteGBSGA{}, and \MSBipartiteGBSGA{}) share common hyperparameters where applicable, including population size $N_{\mathrm{pop}}=100$, maximum generations $G_{\max}=200$ (except \MSBipartiteGBSGA{} which uses $G_{\max}^{\mathrm{MS}}=300$; see below), crossover rate $p_c=0.8$, and mutation rate $p_m=0.1$. BipartiteGBS-specific parameters ($N_{\text{BipartiteGBS}}$, $\eta$, $\alpha_{\max}$, $\beta$) are listed separately. The full set of hyperparameters is given in Tab.~\ref{tab:hyperparams} of Appx.~\ref{appx:hyperparameters}. NN, as a greedy heuristic, uses none of the GA hyperparameters. 

For Erd\H{o}s--R\'enyi graphs with $p=0.3$, a Hamiltonian cycle exists with high probability. However, the search space grows factorially as $n!$, making exhaustive enumeration infeasible for $n\geq 30$. In the standard genetic algorithm, the number of candidate solutions evaluated is limited to $N_{\mathrm{pop}}G_{\max}$, which is orders of magnitude smaller than $n!$. For \MSBipartiteGBSGA{}, we set $G_{\max}^{\mathrm{MS}}=300$ and allocate the budget as: Stage~1 runs for $G_{\max}^{\mathrm{MS}}/3=100$ generations with $N_{\mathrm{pop}}/2=50$ individuals, and Stage~2 runs for $G_{\max}^{\mathrm{MS}}/2=150$ generations with $N_{\mathrm{pop}}=100$ individuals. The total evaluations ($50\times100 + 100\times150 = 20\,000$) equal the baseline GA budget $N_{\mathrm{pop}}G_{\max}=100\times200$. The additional cost of generating BipartiteGBS samples and estimating co-occurrence frequencies is a one-time preprocessing step per graph instance, which is not included in the per-generation comparison.

\subsection{Results and analysis}

\begin{figure*}[t]
\centering
\includegraphics[width=1.0\linewidth]{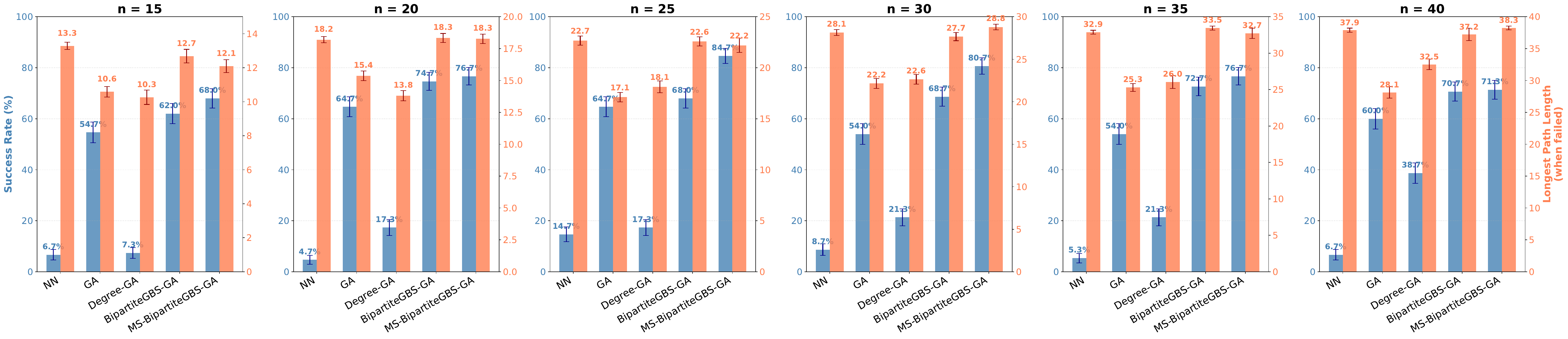}
\caption{\textbf{Main benchmark metrics for Hamiltonian-cycle search.} The compared algorithms are NN, GA, Degree-GA, \BipartiteGBSGA{}, and \MSBipartiteGBSGA{}. Results are shown for Erd\H{o}s--R\'enyi random graphs with edge probability $p=0.3$ and graph sizes $n=15,20,25,30,35,40$. For each size, we generated $30$ graph instances and performed $5$ independent runs per instance, giving $150$ runs for each algorithm. The blue bars, read against the left axis, report the mean success rate, i.e., the fraction of runs that found a valid Hamiltonian cycle. The orange bars, read against the right axis, report the mean longest-path length among failed runs only; this length is the number of vertices in the longest contiguous segment of consecutive valid directed edges in a candidate permutation, ranging from $1$ to $n$ (see \textsc{ExtractLongestPath} in Alg.~\ref{alg:gbs_helper}). For each graph instance, the longest-path length is first averaged over its failed runs, and these per-instance means are then averaged over the $30$ instances. Error bars indicate $\pm$ one standard error of the mean, using binomial standard errors for success rates and standard errors of per-instance means for longest-path lengths.}%
\label{fig:algorithm comparition}
\end{figure*}

Fig.~\ref{fig:algorithm comparition} shows that the BipartiteGBS-enhanced algorithms (\BipartiteGBSGA{} and \MSBipartiteGBSGA{}) consistently achieve higher success rates than the standard GA across all graph sizes. Furthermore, in cases where no Hamiltonian cycle is found, the BipartiteGBS-enhanced methods produce longer valid paths, indicating an improved ability to extract and exploit meaningful partial structural information. The proposed method is a BipartiteGBS-assisted heuristic: its numerical performance suggests that permanent-biased sampling can provide useful guidance for certain directed Hamiltonian-cycle instances, but we do not claim a complexity-theoretic quantum advantage for Hamiltonian-cycle search.

The same figure shows that Degree-GA underperforms even the standard GA, indicating that simple vertex-level degree information does not explain the improvement achieved by BipartiteGBS-guided search. In contrast, BipartiteGBS guidance uses sampled subgraphs and directed-edge co-occurrence frequencies, which provide structural information beyond individual vertex degrees and align more naturally with Hamiltonian cycle search.

\begin{figure*}[t]
\centering
\includegraphics[width=1.0\linewidth]{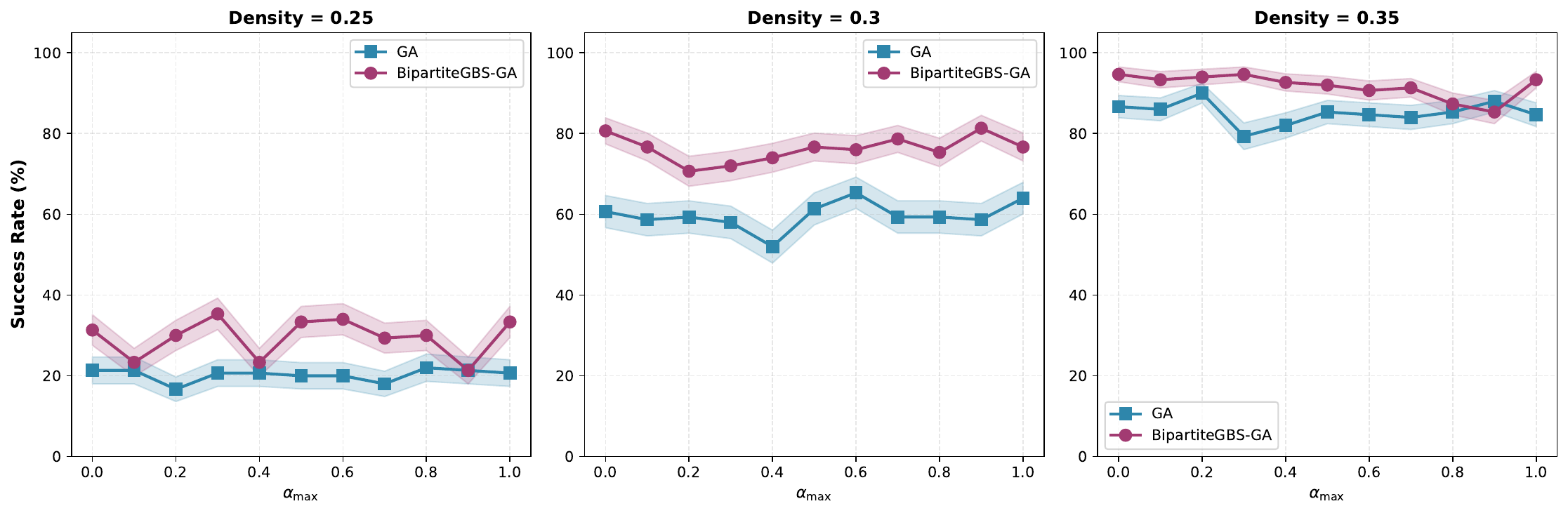}
\caption{\textbf{Parameter-scan setup for the BipartiteGBS fitness weight.} The parameter $\alpha_{\max}$ controls how strongly edge co-occurrence frequencies influence the fitness function relative to raw cycle length. Results are shown for Erd\H{o}s--R\'enyi random graphs with $n=25$ and densities $p = 0.25, 0.30, 0.35$. For each density, we generated $30$ graph instances and performed $5$ independent runs for each value of $\alpha_{\max}$ from $0.0$ to $1.0$ in increments of $0.1$. Curves show mean success rates across all instances, and shaded bands indicate $\pm$ one standard error.}%
\label{fig:algorithm alpha}
\end{figure*}

\textbf{Hyperparameter sensitivity.}
Genetic algorithms introduce multiple hyperparameters that critically influence search behavior and solution quality. To investigate the sensitivity of \BipartiteGBSGA{} to the hyperparameter $\alpha_{\max}$, which balances BipartiteGBS-derived co-occurrence frequencies against raw cycle length in fitness evaluation, we conduct experiments for $\alpha_{\max} \in [0.0,1.0]$ in $0.1$ increments on Erd\H{o}s--R\'enyi random graphs ($n=25$, densities $0.25,0.30,0.35$).
As shown in Fig.~\ref{fig:algorithm alpha}, for all $\alpha_{\max}$, \BipartiteGBSGA{} achieves success rates comparable to or slightly higher than standard GA in most cases, with little apparent sensitivity to the choice of $\alpha_{\max}$. 

\begin{table*}[htbp]
\centering
\setlength{\tabcolsep}{0pt} 
\setlength\heavyrulewidth{0.3ex}  
\renewcommand{\arraystretch}{1.5} 
\begin{tabular}{@{}w{c}{0.18\textwidth}w{c}{0.19\textwidth}w{c}{0.19\textwidth}w{c}{0.19\textwidth}w{c}{0.20\textwidth}@{}}
\toprule
\textbf{Variant} & \textbf{Init} & \textbf{Fitness} & \textbf{Mutation} & \textbf{Info Type} \\
\midrule
GA               & Random & Standard & Random & -- \\
InitOnly         & BipartiteGBS & Standard & Random & Subgraph + Edge \\
FitnessOnly      & Random & BipartiteGBS & Random & Edge only \\
MutationOnly     & Random & Standard & BipartiteGBS & Edge only \\
SubgraphOnly     & BipartiteGBS & Standard & Random & Subgraph only \\
EdgeOnly         & BipartiteGBS & Standard & Random & Edge only \\
\BipartiteGBSGA{} & BipartiteGBS & BipartiteGBS & BipartiteGBS & Subgraph + Edge \\
\MSBipartiteGBSGA{} & \multicolumn{4}{c}{Two-stage adaptive} \\
\bottomrule
\end{tabular}
\caption{\textbf{Definition of algorithmic variants used in the ablation study.} 
The columns specify where BipartiteGBS-derived information is inserted into the genetic algorithm: initialization, fitness evaluation, and mutation. ``Subgraph'' denotes accepted BipartiteGBS sampled vertex subsets, while ``Edge'' denotes directed-edge co-occurrence frequencies estimated from those samples. 
GA is the no-BipartiteGBS baseline, 
\BipartiteGBSGA{} activates all three BipartiteGBS-guided components, and 
\MSBipartiteGBSGA{} adds the two-stage adaptive procedure described in Sec.~\ref{sec:multi-stage-variant}.}
\label{tab:ablation_variants}
\end{table*}

\begin{figure*}[t]
\centering
\includegraphics[width=1.0\linewidth]{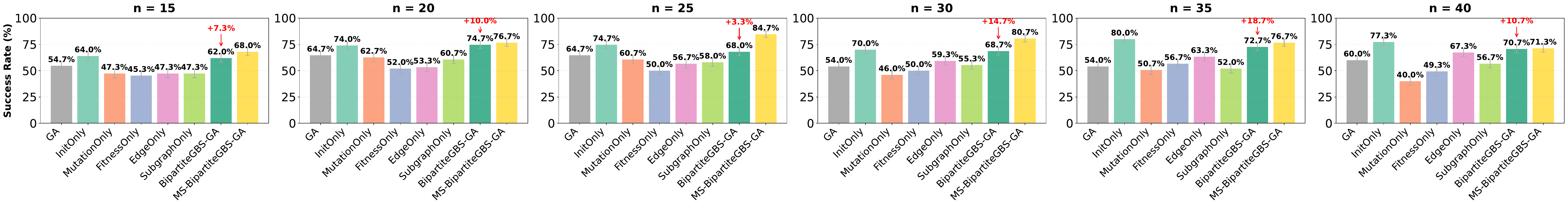}
\caption{\textbf{Ablation study setup and plotted success-rate quantities.}
The compared variants are defined in Tab.~\ref{tab:ablation_variants}. Results are shown for Erd\H{o}s--R\'enyi random directed graphs with edge probability $p=0.3$ and sizes $n=15,20,25,30,35,40$. For each size, $30$ graph instances were generated, and each algorithm was run $5$ times on each instance. Bars show mean success rates over these runs, and error bars indicate $\pm$ one standard error of the mean. Red annotations, where shown, indicate the absolute success-rate difference between \BipartiteGBSGA{} and the no-BipartiteGBS GA baseline in the corresponding subplot; they are not additional error bars.}
\label{fig:ablation}
\end{figure*}

\subsection{Ablation study}\label{sec:ablation-study}

To systematically evaluate the contribution of each BipartiteGBS component, we compared the standard GA baseline, five component-level variants, the \BipartiteGBSGA{}, and \MSBipartiteGBSGA{}. Tab.~\ref{tab:ablation_variants} summarizes how each variant incorporates BipartiteGBS-derived information into initialization, fitness evaluation, and mutation. All component-level variants use the same permutation representation, selection rule, crossover operator, and classical search budget. InitOnly uses both accepted BipartiteGBS subgraphs and edge co-occurrence frequencies only during initialization, while retaining standard fitness and random mutation. FitnessOnly uses edge co-occurrence frequencies only in the fitness function, and MutationOnly uses them only in the mutation operator. SubgraphOnly and EdgeOnly isolate the two information channels in initialization: the former uses only accepted sampled subgraphs, whereas the latter uses only directed-edge co-occurrence frequencies. \BipartiteGBSGA{} activates BipartiteGBS-guided initialization, fitness, and mutation, while \MSBipartiteGBSGA{} further adds the two-stage adaptive procedure. For the EdgeOnly and SubgraphOnly variants, where BipartiteGBS information is limited to a single channel, we adopted a mixed initialization scheme: half of the population via random permutation, half via BipartiteGBS guidance.

As shown in Fig.~\ref{fig:ablation}, InitOnly achieves the highest success rate among the compared variants. In the terminology of Tab.~\ref{tab:ablation_variants}, InitOnly uses BipartiteGBS information only in initialization, with both accepted sampled subgraphs and directed-edge co-occurrence frequencies, while retaining standard fitness and random mutation. The comparison with SubgraphOnly and EdgeOnly shows that both information channels contribute useful initialization guidance, with their combination giving the strongest component-level performance.
In contrast, FitnessOnly and MutationOnly do not provide comparable gains. Full \BipartiteGBSGA{}, which activates BipartiteGBS-guided initialization, fitness, and mutation, underperforms relative to InitOnly, and \MSBipartiteGBSGA{} does not consistently outperform InitOnly. Numerical results suggest that BipartiteGBS-guided fitness and BipartiteGBS-guided mutation may have neutral or slightly negative effects: fitness may introduce biased evaluation signals, while mutation may disrupt beneficial genetic structures.

These findings identify the most useful role of BipartiteGBS in the hybrid search: it acts as a structure-aware proposal mechanism that supplies the classical genetic algorithm with promising initial vertex orderings. Once this initial population is formed, the standard evolutionary operators are better suited to refine and recombine candidate cycles. In contrast, using BipartiteGBS information to continually bias fitness evaluation or mutation can overconstrain the search and interfere with useful genetic diversity. For Hamiltonian cycle search, the most effective design is therefore not the most heavily hybridized one, but the simpler InitOnly strategy, which uses BipartiteGBS to initialize the search and then lets the classical genetic algorithm perform the subsequent optimization.

\section{Conclusions}
\label{sec:conclusions}

In this work, we introduced the Max-Perm problem as a conceptual optimization task for BipartiteGBS and derived a closed-form enhancement factor for an idealized sampling setting, illustrating a theoretical advantage over uniform classical sampling. We then proposed a BipartiteGBS-enhanced genetic algorithm for the Hamiltonian cycle problem in directed graphs. Unlike previous GBS-based graph algorithms that were restricted to undirected graphs, our approach exploited BipartiteGBS to generate samples with probabilities proportional to permanents of arbitrary complex submatrices. Through numerical simulations on Erd\H{o}s--R\'enyi random graphs, we demonstrated that the BipartiteGBS-enhanced genetic algorithms consistently outperformed a standard genetic algorithm in both success rate and the length of the longest valid path when no Hamiltonian cycle was found. In the main comparison, \MSBipartiteGBSGA{}, which first performed a fast classical search and then invoked BipartiteGBS guidance, achieved the best performance among the compared algorithms across the tested graph sizes. In the ablation study, however, InitOnly achieved the highest success rate among the component-level variants, and \MSBipartiteGBSGA{} did not consistently outperform InitOnly. Our ablation study further revealed an important design principle: the most substantial contribution of BipartiteGBS came from BipartiteGBS-guided initialization, while incorporating BipartiteGBS information into fitness evaluation and mutation operators yielded diminishing or neutral returns.

Two problems deserve further research. First, an experimental implementation on photonic chips is needed to evaluate how loss, finite sampling, imperfect squeezing, and device constraints affect the usefulness of BipartiteGBS guidance in practical directed-graph search. Second, the framework naturally extends to other directed graph problems, such as the traveling salesman problem, where permanent-biased sampling may provide useful structural information.

\section*{Acknowledgements}

This work was supported by 
the National Key R\&D Program of China (Grant No. 2022YFF0712800),
the Quantum Science and Technology-National Science and Technology Major Project (Grant No. 2025ZD0300300) and 
the National Natural Science Foundation of China (Grant No. 12504584).

\section*{Data and Code Availability}

The numerical data and Python code used to generate the tables and figures in this work are available
upon reasonable request to the corresponding authors.

%

\makeatletter
\newcommand{\appendixtitle}[1]{\gdef\@title{#1}}
\makeatother

\makeatletter%
\newcommand{\appendixmaketitle}{%
\begin{center}%
\vspace{0.4in}%
{\large \@title \par}%
\end{center}%
\par%
}%
\makeatother%

\makeatletter
\newcommand{\appendixtableofcontents}{%
\begingroup
\setcounter{tocdepth}{4}%
\@starttoc{atoc}%
\endgroup}
\makeatother

\newcommand{\appendixtocdivider}{%
\par\medskip
\noindent\hbox to \linewidth{%
\leaders\hbox{\rule[0.6ex]{1pt}{0.4pt}}\hfill
\hspace{0.8em}\textsc{Table of Contents}\hspace{0.8em}%
\leaders\hbox{\rule[0.6ex]{1pt}{0.4pt}}\hfill}%
\par}

\makeatletter
\newcounter{subsubsubsection}[subsubsection]
\renewcommand{\thesubsubsubsection}{\thesubsubsection.\arabic{subsubsubsection}}
\providecommand{\theHsubsubsubsection}{}
\renewcommand{\theHsubsubsubsection}{\theHsubsubsection.\arabic{subsubsubsection}}
\providecommand*{\toclevel@subsubsubsection}{4}
\newcommand{\subsubsubsection}[1]{%
\refstepcounter{subsubsubsection}%
\paragraph*{\thesubsubsubsection\space #1}}
\newcommand*\l@subsubsubsection[2]{\@dottedtocline{4}{4.4em}{3.0em}{#1}{#2}}
\makeatother

\newcommand{\appsection}[1]{%
\section{#1}%
\addcontentsline{atoc}{section}{\protect\numberline{\thesection}#1}}

\newcommand{\appsubsection}[1]{%
\subsection{#1}%
\addcontentsline{atoc}{subsection}{\protect\numberline{\thesubsection}#1}}

\newcommand{\appsubsubsection}[1]{%
\subsubsection{#1}%
\addcontentsline{atoc}{subsubsection}{\protect\numberline{\thesubsubsection}#1}}

\newcommand{\appsubsubsubsection}[1]{%
\subsubsubsection{#1}}

\setcounter{secnumdepth}{4}
\appendix
\widetext
\newpage

\appendixtitle{\bf
Supplemental Material for\\``\thetitle''}
\appendixmaketitle
\vspace{0.1in}

This Supplemental Material is organized as follows. Appx.~\ref{appx:BipartiteGBS-GA} gives the full specification of the BipartiteGBS-enhanced genetic algorithm (\BipartiteGBSGA{}), including the BipartiteGBS sampler, evolution loop, initialization procedure, fitness evaluation, mutation operator, helper routines, and hyperparameter settings.
Appx.~\ref{appx:MS-BipartiteGBS-GA} describes \MSBipartiteGBSGA{}.
Appx.~\ref{appx:classical-benchmarks} collects the three classical benchmark algorithms used for comparison: the nearest neighbor heuristic, the standard genetic algorithm, and the degree-centrality-enhanced genetic algorithm.

\appendixtocdivider
{%
\appendixtableofcontents%
}%

\appsection{Details for the \BipartiteGBSGA{} algorithm}\label{appx:BipartiteGBS-GA}

This section presents the full algorithmic specification of the \BipartiteGBSGA{} framework used in the numerical experiments, including the BipartiteGBS sampling procedure, the genetic-algorithm workflow, the BipartiteGBS-guided initialization, the BipartiteGBS-enhanced fitness, and the BipartiteGBS-guided mutation components, the helper routines, and the hyperparameter settings.

\appsubsection{\BipartiteGBSGA{} algorithm}

The overall structure and call relationships are as follows:

\begin{itemize}
    \item \textbf{BipartiteGBS sampler (Alg.~\ref{alg:gbs_sampler_simple}).}
    Performs SVD on the adjacency matrix of a directed graph, encodes it into BipartiteGBS device parameters, draws samples, and produces co-occurrence frequencies and projected vertex subsets. Only collision-free patterns with $|S|=|T|\ge 3$ are accepted.
    \textbf{Outputs}: co-occurrence frequencies $p_{ed}(u,v)$ and the set of projected vertex subsets $\mathcal{S}$.

    \item \textbf{\BipartiteGBSGA{} overall procedure (Alg.~\ref{alg:gbs_ga}).}
    Top-level driver that calls the BipartiteGBS sampler to obtain co-occurrence frequencies and vertex subsets, then launches the genetic search.

    \item \textbf{\BipartiteGBSGA{} evolution loop (Alg.~\ref{alg:gbs_ga_main}).}
    Runs the population-based genetic search using the BipartiteGBS information.
    \textbf{Calls}: Alg.~\ref{alg:gbs_init} for initialization, Alg.~\ref{alg:gbs_fitness} for fitness evaluation, Alg.~\ref{alg:gbs_mutation} for mutation, and Alg.~\ref{alg:gbs_helper} for standard genetic helper functions.

    \item \textbf{BipartiteGBS-guided initialization (Alg.~\ref{alg:gbs_init}).}
    Hybrid initialization with three strategies: random permutations, paths from BipartiteGBS samples (calls Alg.~\ref{alg:buildpath}), and hybrid construction (calls Alg.~\ref{alg:hybridpath}).

    \item \textbf{BuildPath (Alg.~\ref{alg:buildpath}).}
    Given a projected vertex subset $S$ from BipartiteGBS sampling, greedily connects edges in descending order of co-occurrence frequency to construct a path.

    \item \textbf{HybridPath (Alg.~\ref{alg:hybridpath}).}
    Hybrid path construction that probabilistically uses BipartiteGBS top edges and random exploration.

    \item \textbf{BipartiteGBS-enhanced fitness (Alg.~\ref{alg:gbs_fitness}).}
    Computes fitness by combining the number of valid edges (base fitness) with the average co-occurrence frequency of those edges.

    \item \textbf{BipartiteGBS-guided mutation (Alg.~\ref{alg:gbs_mutation}).}
    Identifies the edge with the lowest co-occurrence frequency in the cycle and swaps one of its endpoints with a random vertex.

    \item \textbf{Helper functions (Alg.~\ref{alg:gbs_helper}).}
    Contains \textsc{IsHamiltonianCycle}, \textsc{TournamentSelect}, and \textsc{OrderCrossover}.

\end{itemize}

\vspace{0.2in}
\hrule
\vspace{0.2in}

\begin{algorithm}[H]
\caption{BipartiteGBS Sampler}
\label{alg:gbs_sampler_simple}
\DontPrintSemicolon
\SetKwInOut{Input}{Input}
\SetKwInOut{Output}{Output}

\Input{
    Directed graph $G=(V,E)$, $|V|=n$ \\
    Adjacency matrix $A$ \\
    Number of samples $N$
}
\Output{
    Co-occurrence frequencies $p_{ed}(u,v)$, projected vertex subsets $\mathcal{S}$
}

\Begin{
    \tcc{Encode graph into BipartiteGBS device}
    $U, \Sigma, V^\dagger \leftarrow \operatorname{SVD}(A)$ \;
    $\{r_i\} \leftarrow \operatorname{arctanh}(\eta \cdot \Sigma_{ii} / \max_j \Sigma_{jj})$ \;
    $\mathcal{S} \leftarrow \emptyset$, $\text{counts} \leftarrow \{\}$ \;

    \For{$k \leftarrow 1$ \KwTo $N$}{
        \tcc{Sample from BipartiteGBS}
        $(S,T) \leftarrow \textsc{BipartiteGBS}(U, V, \{r_i\})$ \;
        \tcc{Accept only collision-free patterns with balanced photons}
        \If{$|S| = |T| \ge 3$ \textbf{and} no photon number $>1$}{
            \tcc{Heuristically project the row/source and column/target registers back to one vertex pool}
            $V_S \leftarrow S \cup T$ \;
            \tcc{$V_S$ is a projected candidate vertex pool, not the sampled submatrix $A_{S,T}$ itself}
            $\mathcal{S} \leftarrow \mathcal{S}.append(V_S)$ \;

            \ForEach{$(u,v) \in V_S \times V_S,\ u \neq v$}{
                \If{$A_{uv}=1$}{$\text{counts}[(u,v)] \leftarrow \text{counts}[(u,v)] + 1$}
            }
        }
    }
    $M \leftarrow \max(|\mathcal{S}|, 1)$\;
    $p_{ed}(u,v) \leftarrow \text{counts}[(u,v)] / M$

    \Return $p_{ed}(u,v)$, $\mathcal{S}$ \;
}
\end{algorithm}

\begin{algorithm}[H]
\caption{BipartiteGBS-Enhanced Genetic Algorithm (\BipartiteGBSGA{})}
\label{alg:gbs_ga}
\DontPrintSemicolon
\SetKwInOut{Input}{Input}
\SetKwInOut{Output}{Output}

\Input{
    Directed graph $G=(V,E)$, $|V|=n$ \\
    Adjacency matrix $A$ \\
    Number of BipartiteGBS samples $N_{\text{BipartiteGBS}}$ \\
    Population size $N_{\text{pop}}$, maximum generations $G_{\max}$ \\
    Crossover rate $p_c$, mutation rate $p_m$ \\
    Maximum fitness weight $\alpha_{\max}$, hybrid parameter $\beta$
}
\Output{
    Hamiltonian cycle if found; otherwise the longest path found
}

\Begin{
    \tcc{Step 1: Extract BipartiteGBS guidance from the directed graph}
    $(p_{ed}, \mathcal{S}) \leftarrow \textsc{BipartiteGBSSampler}(G, A, N_{\text{BipartiteGBS}})$ (Alg.~\ref{alg:gbs_sampler_simple}) \;

    \tcc{Step 2: Run the genetic search using the BipartiteGBS guidance}
    $result \leftarrow \textsc{BipartiteGBSGAEvolutionLoop}(G, p_{ed}, \mathcal{S}, N_{\text{pop}}, G_{\max}, p_c, p_m, \alpha_{\max}, \beta)$ (Alg.~\ref{alg:gbs_ga_main}) \;

    \Return $result$ \;
}
\end{algorithm}

\begin{algorithm}[H]
\caption{\BipartiteGBSGA{} Evolution Loop}
\label{alg:gbs_ga_main}
\DontPrintSemicolon
\SetKwInOut{Input}{Input}
\SetKwInOut{Output}{Output}

\Input{
    Graph $G=(V,E)$, $|V|=n$ \\
    BipartiteGBS information: co-occurrence frequencies $p_{ed}(u,v)$, samples $\mathcal{S}$ \\
    Population size $N_{\text{pop}}$, max generations $G_{\max}$ \\
    Crossover rate $p_c$, mutation rate $p_m$, weight $\alpha_{\max}$, hybrid parameter $\beta$
}
\Output{
    Best found Hamiltonian cycle or longest path
}

\Begin{
    \tcc{Step 1: Initialize population}
    $P \leftarrow \textsc{BipartiteGBSGuidedInitialization}(G, \mathcal{S}, p_{ed}, N_{\text{pop}}, \beta)$ (Alg.~\ref{alg:gbs_init}) \;
    
    \tcc{Step 2: Track best individual across all generations}
    $best \leftarrow \text{empty}$ \;
    $bestFitness \leftarrow 0$ \;
    $bestCycle \leftarrow \text{empty}$ \;
    
    \tcc{Step 3: Main evolution loop}
    \For{$gen \leftarrow 1$ \KwTo $G_{\max}$}{
        $\alpha_{\text{cur}} \leftarrow \alpha_{\max} \cdot gen / G_{\max}$ \;
        
        \tcc{Evaluate fitness for each individual}
        \ForEach{$ind \in P$}{
            $fitness[ind] \leftarrow \textsc{BipartiteGBSFitness}(G, ind, p_{ed}, \alpha_{\text{cur}})$ (Alg.~\ref{alg:gbs_fitness}) \;
        }
        
        \tcc{Update best-so-far individual}
        $curBest \leftarrow \arg\max fitness$ \;
        \If{$fitness[curBest] > bestFitness$}{
            $bestFitness \leftarrow fitness[curBest]$ \;
            $best \leftarrow curBest$ \;
            \If{$\textsc{IsHamiltonianCycle}(best, G)$ (Alg.~\ref{alg:gbs_helper})}{
                $bestCycle \leftarrow best$ \;
                \Return $bestCycle$ \tcc*[r]{Found Hamiltonian cycle}
            }
        }
        
        \tcc{Generate next population}
        $P_{\text{new}} \leftarrow \emptyset$ \;
        \For{$i \leftarrow 1$ \KwTo $N_{\text{pop}}$}{
            $p_1 \leftarrow \textsc{TournamentSelect}(P, fitness)$ (Alg.~\ref{alg:gbs_helper}) \;
            $p_2 \leftarrow \textsc{TournamentSelect}(P, fitness)$ (Alg.~\ref{alg:gbs_helper}) \;
            \If{$\text{random}() < p_c$}{
                $child \leftarrow \textsc{OrderCrossover}(p_1, p_2)$ (Alg.~\ref{alg:gbs_helper}) \;
            }
            \Else{
                $child \leftarrow p_1$ \;
            }
            \If{$\text{random}() < p_m$}{
                $child \leftarrow \textsc{BipartiteGBSMutate}(child, p_{ed})$ (Alg.~\ref{alg:gbs_mutation}) \;
            }
            
            $P_{\text{new}}.\text{append}(child)$ \;
        }
        $P \leftarrow P_{\text{new}}$ \;
    }
    
    \tcc{Return best result found (not just last generation)}
    \If{$bestCycle \neq \text{empty}$}{
        \Return $bestCycle$ \;
    }
    \Else{
        $longestPath \leftarrow \textsc{ExtractLongestPath}(G, best)$ (Alg.~\ref{alg:gbs_helper}) \;
        \Return $longestPath$ \;
    }
}

\end{algorithm}

\begin{algorithm}[H]
\caption{BipartiteGBS-Guided Initialization}
\label{alg:gbs_init}
\DontPrintSemicolon
\SetKwInOut{Input}{Input}
\SetKwInOut{Output}{Output}

\Input{
    Graph $G=(V,E)$, $|V|=n$ \\
    BipartiteGBS information: co-occurrence frequencies $p_{ed}(u,v)$, samples $\mathcal{S}$ \\
    Population size $N$, hybrid parameter $\beta$ (default: 0.5)
}
\Output{
    Initial population $P$ (size $N$)
}

\Begin{
    $P \leftarrow \text{empty list}$ \;
    
    \tcc{Strategy 1: Random permutations (33.3\%)}
    \For{$i \leftarrow 1$ \KwTo $N/3$}{
        $ind \leftarrow \text{random permutation of } [0, \ldots, n-1]$ \;
        $P.\text{append}(ind)$ \;
    }
    
    \tcc{Strategy 2: BipartiteGBS-guided construction from subgraphs (33.3\%)}
    \If{$\mathcal{S} \neq \emptyset$}{
        $k \leftarrow \min(N/3, |\mathcal{S}|)$ \;
        \For{$i \leftarrow 1$ \KwTo $k$}{
            $sample \leftarrow \mathcal{S}[i-1]$ \;
            $ind \leftarrow \textsc{BuildPath}(G, sample, p_{ed})$ (Alg.~\ref{alg:buildpath}) \;
            \If{$ind \neq \text{NULL}$}{
                $P.\text{append}(ind)$ \;
            }
        }
    }
    
    \tcc{Strategy 3: Hybrid construction (33.3\%)}
    \For{$i \leftarrow 1$ \KwTo $N/3$}{
        $ind \leftarrow \textsc{HybridPath}(G, p_{ed}, \mathcal{S}, \beta)$ (Alg.~\ref{alg:hybridpath}) \;
        \If{$ind \neq \text{NULL}$}{
            $P.\text{append}(ind)$ \;
        }
    }
    
    \tcc{Fill remaining slots with random permutations}
    \While{$|P| < N$}{
        $ind \leftarrow \text{random permutation of } [0, \ldots, n-1]$ \;
        $P.\text{append}(ind)$ \;
    }
    
    \Return $P[0:N]$ \;

}
\end{algorithm}

\begin{algorithm}[H]
\caption{BuildPath: Construct Path from BipartiteGBS information}
\label{alg:buildpath}
\DontPrintSemicolon
\SetKwInOut{Input}{Input}
\SetKwInOut{Output}{Output}

\Input{
    Graph $G=(V,E)$, vertex set $S$ from a BipartiteGBS sample, co-occurrence frequencies $p_{ed}(u,v)$
}
\Output{
    Individual path $ind$ or NULL
}

\Begin{
    $\text{vertices} \leftarrow \text{sorted}(S)$ \;
    \If{$|\text{vertices}| < 3$}{\Return NULL}
    
    $\text{used} \leftarrow \emptyset$, $\text{path} \leftarrow []$ \;
    $\text{start} \leftarrow \text{vertices}[0]$ \;
    $\text{path}.\text{append}(\text{start})$, $\text{used}.\text{add}(\text{start})$ \;
    $\text{current} \leftarrow \text{start}$, $\text{remaining} \leftarrow \text{vertices}[1:]$ \;
    
    \While{$\text{remaining} \neq \emptyset$}{
        $\text{candidates} \leftarrow []$ \;
        \ForEach{$v \in \text{remaining}$}{
            \If{$G.\text{has\_edge}(\text{current}, v)$}{

                \tcc{Query co-occurrence frequency; default 0.5 for unsampled edges}
                $p(current,v) \leftarrow p_{ed}((current,v), 0.5)$ \;
                $\text{candidates}.\text{append}((v, p(\text{current}, v)))$ \;
            }
        }
        \If{$\text{candidates} \neq \emptyset$}{
            sort $\text{candidates}$ by probability descending \;
            $\text{next} \leftarrow \text{candidates}[0][0]$ \;
            $\text{path}.\text{append}(\text{next})$, $\text{used}.\text{add}(\text{next})$ \;
            $\text{current} \leftarrow \text{next}$, $\text{remaining}.\text{remove}(\text{next})$ \;
        }
        \Else{
            \textbf{break} \;
        }
    }
    
    \tcc{Append all unused graph vertices in random order}
    $remaining2 \leftarrow V \setminus used$ \;
    $remaining2 \leftarrow \text{random shuffle of } remaining2$ \;
    $path.\text{extend}(remaining2)$ \;

    \Return $\text{path}$ \;
}
\end{algorithm}

\begin{algorithm}[H]
\caption{HybridPath: Hybrid Path Construction Using BipartiteGBS information}
\label{alg:hybridpath}
\DontPrintSemicolon
\SetKwInOut{Input}{Input}
\SetKwInOut{Output}{Output}

\Input{
    Graph $G=(V,E)$, $|V|=n$ \\
    Co-occurrence frequencies $p_{ed}(u,v)$ \\
    Sampled subgraphs $\mathcal{S}$, hybrid parameter $\beta$
}
\Output{
    Individual path $ind$
}

\Begin{

    $\text{used} \leftarrow \emptyset$, $\text{path} \leftarrow []$ \;
    
    \tcc{With 50\% probability, use BipartiteGBS information to choose start vertex}
    \uIf{$\text{random}() < 0.5$ and $\mathcal{S} \neq \emptyset$}{
        \tcc{Randomly pick one BipartiteGBS-sampled subgraph (from first 50 samples)}
        $\text{sample} \leftarrow \text{random.choice}(\mathcal{S}[:50])$ \;
        \tcc{Randomly pick a vertex from that subgraph as the starting point}
        $\text{start} \leftarrow \text{random.choice}(\text{sample})$ \;
    }
    \Else{

        $\text{start} \leftarrow \text{random}(0, n-1)$ \;
    }
    
    $\text{path}.\text{append}(\text{start})$, $\text{used}.\text{add}(\text{start})$ \;
    
    \tcc{Keep building until the path contains all $n$ vertices}

   $k \leftarrow \min(2 \cdot n, |p_{ed}|)$ \;
   $E_{\text{top}} \leftarrow \text{top } k \text{ directed edges sorted by } p_{ed}(u,v) \text{ in descending order}$ \;
    
    \While{$|\text{path}| < n$}{
        \tcc{Get the last vertex in the current path}
        $\text{last} \leftarrow \text{path}[-1]$ \;
        
        \tcc{With probability $\beta$, try to use BipartiteGBS top edges; if no match, fall through to random exploration}
        $found \leftarrow \text{false}$ \;
        \If{$\text{random}() < \beta$ and $E_{\text{top}} \neq \emptyset$}{
            \ForEach{$(u,v) \in E_{\text{top}}[:n]$}{
                \If{$u = \text{last}$ and $v \notin \text{used}$}{
                    $\text{path}.\text{append}(v)$, $\text{used}.\text{add}(v)$ \;
                    $found \leftarrow \text{true}$ \;
                    \textbf{break} \;
                }
            }
        }
        \If{$\neg found$}{
            \tcc{Fall back to random exploration}
            $\text{neighbors} \leftarrow \{v : G.\text{has\_edge}(\text{last}, v) \text{ and } v \notin \text{used}\}$ \;
            \If{$\text{neighbors} \neq \emptyset$}{
                
                $\text{next} \leftarrow \text{random.choice}(\text{neighbors})$ \;
                
                $\text{path}.\text{append}(\text{next})$, $\text{used}.\text{add}(\text{next})$ \;
            }
            \Else{
                \tcc{If no valid neighbor exists from current vertex}
                \tcc{Find all vertices not yet visited}
                $\text{remaining} \leftarrow V \setminus \text{used}$ \;
                \If{$\text{remaining} \neq \emptyset$}{
                    \tcc{If there are still unvisited vertices, randomly pick one and append}
                    $\text{path}.\text{append}(\text{random.choice}(\text{remaining}))$ \;
                    $\text{used}.\text{add}(\text{path}[-1])$ \;
                }
                \Else{
                    \textbf{break} \;
                }
            }
        }
    }
    
    \tcc{Ensure all vertices are included (safety net for edge cases)}
    $\text{remaining} \leftarrow V \setminus \text{used}$ \;
    
    $\text{path}.\text{extend}(\text{remaining})$ \;
    
    \Return $\text{path}$ \;
}
\end{algorithm}

\begin{algorithm}[H]
\caption{BipartiteGBS-Enhanced Fitness}
\label{alg:gbs_fitness}
\DontPrintSemicolon
\SetKwInOut{Input}{Input}
\SetKwInOut{Output}{Output}

\Input{
    Graph $G=(V,E)$, $|V|=n$ \\
    Individual $ind=[v_0,\ldots,v_{n-1}]$ (a permutation of vertices) \\
    Co-occurrence frequencies $p_{ed}(u,v)$ from BipartiteGBS sampling \\
    Weight $\alpha_{\text{cur}} \in [0,1]$ (increases with generation)
}
\Output{
    Fitness value $f \in [0,1]$
}

\Begin{
    \tcc{Step 1: Count the number of valid edges in the path (base fitness)}
    $len \leftarrow 0$ \;
    \For{$i=0$ \KwTo $n-2$}{
        \If{$G.\text{has\_edge}(ind[i], ind[i+1])$}{
            $len \leftarrow len + 1$ \;
        }
    }
    
    $canClose \leftarrow G.\text{has\_edge}(ind[n-1], ind[0])$ \;
    \If{$canClose$}{
        $len \leftarrow len + 1$ \;
    }
    
    \If{$len = n$}{
        \Return $1.0$ \;
    }
    
    $base \leftarrow len / n$ \;
    
    \tcc{Step 2: Compute average co-occurrence frequency over the same proposed edges used in the implementation}
    $bipartiteScore \leftarrow 0$ \;
    $edgeCount \leftarrow 0$ \;

    \For{$i=0$ \KwTo $n-2$}{
        
        $bipartiteScore \leftarrow bipartiteScore + p_{ed}((ind[i], ind[i+1]), 0)$ \;
        $edgeCount \leftarrow edgeCount + 1$ \;
    }
    \If{$canClose$}{
        $bipartiteScore \leftarrow bipartiteScore + p_{ed}((ind[n-1],ind[0]),0)$ \;
        $edgeCount \leftarrow edgeCount + 1$ \;
    }

    $bipartiteScore \leftarrow bipartiteScore / edgeCount$ \;
    
    \tcc{Step 3: Combine base fitness and BipartiteGBS-derived score}
    \Return $base \times (1 - \alpha_{\text{cur}}) + bipartiteScore \times \alpha_{\text{cur}}$ \;
}
\end{algorithm}

\begin{algorithm}[H]
\caption{BipartiteGBS-Guided Mutation}
\label{alg:gbs_mutation}
\DontPrintSemicolon
\SetKwInOut{Input}{Input}
\SetKwInOut{Output}{Output}

\Input{
    Individual $ind=[v_0,\ldots,v_{n-1}]$ (a permutation of vertices) \\
    Co-occurrence frequencies $p_{ed}(u,v)$
}
\Output{
    Mutated individual $ind'$
}

\Begin{
    $n \leftarrow |ind|$ \;
    
    \tcc{With 70\% probability, use BipartiteGBS information to guide mutation}
    \uIf{$\text{random}() < 0.7$ and $p_{ed} \neq \emptyset$}{
        \tcc{Step 1: Compute BipartiteGBS-derived score for each edge in the cycle}
        $\text{edge\_scores} \leftarrow []$ \;
        \For{$i=0$ \KwTo $n-1$}{
            $u \leftarrow ind[i]$ \;
            $v \leftarrow ind[(i+1) \mod n]$ \;
            \tcc{Query co-occurrence frequency; default 0.5 for unsampled edges}
            $\text{score} \leftarrow p_{ed}((u, v), 0.5)$ \;
            $\text{edge\_scores}.\text{append}((i, \text{score}))$ \;
        }
        
        \tcc{Step 2: Find the edge with the lowest co-occurrence frequency}
        $\text{sort } \text{edge\_scores} \text{ by score ascending}$ \;
        $idx \leftarrow \text{edge\_scores}[0][0]$ \;
        
        \tcc{Step 3: Swap one endpoint of the worst edge with a random vertex}
        $jdx \leftarrow \text{random}(0, n-1)$ \;
        $\text{swap}(ind[idx], ind[jdx])$ \;
        
        \Return $ind$ \;
    }
    \Else{
        \tcc{With 30\% probability, fall back to standard random swap mutation}
        $(i, j) \leftarrow \text{random.sample}([0,\ldots,n-1], 2)$ \;
        $\text{swap}(ind[i], ind[j])$ \;
        \Return $ind$ \;
    }
}
\end{algorithm}

\begin{algorithm}[H]
\caption{Helper Functions}
\label{alg:gbs_helper}
\DontPrintSemicolon
\SetKwInOut{Input}{Input}
\SetKwInOut{Output}{Output}

\BlankLine
\noindent \textbf{Function} IsHamiltonianCycle($ind, G$):
\Begin{
    \tcc{Check: cycle length equals number of vertices}
    \If{$|ind| \neq |V(G)|$}{
        \Return False
    }
    
    \tcc{Check: the individual contains exactly the graph vertices}
    \If{$\text{set}(ind) \neq V(G)$}{
        \Return False
    }
    
    \tcc{Check: consecutive edges exist}
    \For{$i \leftarrow 0$ \KwTo $|ind|-2$}{
        \If{$\textbf{not } G.\text{has\_edge}(ind[i], ind[i+1])$}{
            \Return False
        }
    }
    
    \tcc{Check: closing edge from last to first vertex}
    \Return $G.\text{has\_edge}(ind[-1], ind[0])$
}

\BlankLine
\noindent \textbf{Function} ExtractLongestPath($G, seq$):
\Begin{
    \If{$seq = \emptyset$}{\Return $\emptyset$}
    
    $longest \leftarrow [seq[0]]$ \;
    $current \leftarrow [seq[0]]$ \;
    
    \For{$i \leftarrow 1$ \KwTo $|seq|-1$}{
        \If{$G.\text{has\_edge}(seq[i-1], seq[i])$}{
            $current.\text{append}(seq[i])$ \;
        }
        \Else{
            $longest \leftarrow \arg\max(|longest|, |current|)$ \;
            $current \leftarrow [seq[i]]$ \;
        }
    }
    
    $longest \leftarrow \arg\max(|longest|, |current|)$ \;
    \Return $longest$ \;
}

\BlankLine
\noindent \textbf{Function} TournamentSelect($P, fitness, k=3$):
\Begin{
    \tcc{Randomly select $k$ distinct individuals from the population}
    $\text{indices} \leftarrow \text{random sample of size } \min(k, |P|) \text{ from } \{0, 1, \ldots, |P|-1\}$ \;
    
    \tcc{Find the index with the highest fitness among the selected individuals}
    $\text{best} \leftarrow \arg\max_{i \in \text{indices}} \text{fitness}[i]$ \;
    
    \tcc{Return the individual with the highest fitness}
    \Return $P[\text{best}]$ \;
}

\BlankLine
\noindent \textbf{Function} OrderCrossover($p_1, p_2$):
\Begin{
    \tcc{Step 1: Initialize child with empty slots}
    $n \leftarrow |p_1|$, $child \leftarrow [-1] \times n$ \;
    
    \tcc{Step 2: Randomly select a contiguous segment from parent1}
    $start \leftarrow \text{random}(0, n-2)$, $end \leftarrow \text{random}(start+1, n-1)$ \;
    
    \tcc{Step 3: Copy the selected segment from parent1 to child}
    $child[start:end+1] \leftarrow p_1[start:end+1]$ \;
    
    \tcc{Step 4: Initialize position to fill remaining slots}
    $pos \leftarrow (end+1) \mod n$ \;
    
    \tcc{Step 5: Fill remaining positions with genes from parent2 in cyclic order}
    \For{$i \leftarrow 0$ \KwTo $n-1$}{
        $idx \leftarrow (end+1+i) \mod n$ \;
        \If{$p_2[idx] \notin child$}{
            $child[pos] \leftarrow p_2[idx]$ \;
            $pos \leftarrow (pos+1) \mod n$ \;
        }
    }
    
    \Return $child$ \;
}
\end{algorithm}

\appsubsection{Hyperparameters of the \BipartiteGBSGA{} algorithm}\label{appx:hyperparameters}

The hyperparameters used in the numerical experiments are summarized in Tab.~\ref{tab:hyperparams}. They can be divided into three groups. The first group, including the population size $N_{\text{pop}}$, maximum number of generations $G_{\max}$, crossover rate $p_c$, mutation rate $p_m$, and tournament size $k$, controls the classical genetic-search budget and evolutionary dynamics. These parameters are kept fixed across the GA, Degree-GA, \BipartiteGBSGA{}, and \MSBipartiteGBSGA{} comparisons whenever applicable, so that performance differences are not caused by changing the classical search effort.

\begin{table}[htbp]
\setlength{\tabcolsep}{0pt} 
\setlength\heavyrulewidth{0.3ex}  
\renewcommand{\arraystretch}{1.5} 
\begin{tabular}{@{}w{c}{0.3\textwidth}w{c}{0.3\textwidth}w{c}{0.3\textwidth}@{}}
\toprule
\textbf{Parameter} & \textbf{Symbol} & \textbf{Default value} \\
\midrule
Population size & $N_{\text{pop}}$ & 100 \\
Maximum generations & $G_{\max}$ & 200 \\
Crossover rate & $p_c$ & 0.8 \\
Mutation rate & $p_m$ & 0.1 \\
Tournament size & $k$ & 3 \\
BipartiteGBS hybrid parameter & $\beta$ & 0.2 \\
Fitness weight (max) & $\alpha_{\max}$ & 0.1 \\
Squeezing scaling factor & $\eta$ & 0.75 \\
Number of BipartiteGBS samples & $N_{\text{BipartiteGBS}}$ & 500 \\
\bottomrule
\end{tabular}
\caption{\textbf{Default numerical hyperparameters for the benchmark algorithms.}
The table lists the genetic-algorithm parameters and BipartiteGBS-specific parameters used unless otherwise stated.}
\label{tab:hyperparams}
\end{table}

The second group controls how BipartiteGBS information is incorporated into the genetic algorithm. The hybrid parameter $\beta$ determines how often the initialization procedure attempts to follow edges with high co-occurrence frequencies in the hybrid path-construction step. The maximum fitness weight $\alpha_{\max}$ sets the largest contribution of the co-occurrence frequencies to the fitness function; in the dynamic schedule used in the main comparison, the actual weight increases linearly from zero to $\alpha_{\max}$ over generations. The third group specifies the BipartiteGBS sampling configuration. The squeezing scaling factor $\eta$ rescales the singular values of the graph adjacency matrix before converting them to squeezing parameters, ensuring that the encoded singular values remain below one. The number of BipartiteGBS samples $N_{\text{BipartiteGBS}}$ sets the finite sampling budget used to estimate sampled vertex subsets and empirical co-occurrence frequencies. Unless otherwise stated, all reported experiments use the default values in Tab.~\ref{tab:hyperparams}.

\appsection{Details for the Multi-stage \BipartiteGBSGA{} algorithm}
\label{appx:MS-BipartiteGBS-GA}

\MSBipartiteGBSGA{} uses $N_{\mathrm{pop}}=100$ and a total generation budget $G_{\max}^{\mathrm{MS}}=300$ ($G_{\max}^{\mathrm{MS}}/3=100$ for Stage~1, $G_{\max}^{\mathrm{MS}}/2=150$ for Stage~2), giving $20\,000$ total evaluations—matching the baseline GA budget. It first runs a reduced classical genetic search to solve easy instances cheaply. If this first stage does not find a Hamiltonian cycle, the best partial path found by the classical stage is used to reinforce the BipartiteGBS-derived co-occurrence frequencies before launching the full \BipartiteGBSGA{} evolution loop.

\begin{algorithm}[H]
\caption{Multi-stage \BipartiteGBSGA{} (\MSBipartiteGBSGA{})}
\label{alg:ms_bipartitegbs_ga}
\DontPrintSemicolon
\SetKwInOut{Input}{Input}
\SetKwInOut{Output}{Output}

\Input{
    Directed graph $G=(V,E)$, $|V|=n$ \\
    Adjacency matrix $A$ \\
    Number of BipartiteGBS samples $N_{\text{BipartiteGBS}}$ \\
    Population size $N_{\text{pop}}$, maximum generations $G_{\max}$ \\
    Crossover rate $p_c$, mutation rate $p_m$ \\
    Maximum fitness weight $\alpha_{\max}$, hybrid parameter $\beta$
}
\Output{
    Hamiltonian cycle if found; otherwise the longest path found
}

\Begin{
    \tcc{Stage 1: fast classical search}
    $P_{\text{fast}} \leftarrow \textsc{StandardGA}(G, N_{\text{pop}}/2, G_{\max}/3, p_c, p_m)$ (Alg.~\ref{alg:standard_ga}) \;
    \If{$\textsc{IsHamiltonianCycle}(P_{\text{fast}},G)$ (Alg.~\ref{alg:gbs_helper})}{
        \Return $P_{\text{fast}}$ \;
    }
    
    \tcc{Stage 2: construct BipartiteGBS guidance and reinforce the best classical path}
    $(p_{ed}, \mathcal{S}) \leftarrow \textsc{BipartiteGBSSampler}(G,A,N_{\text{BipartiteGBS}})$ (Alg.~\ref{alg:gbs_sampler_simple}) \;
    \ForEach{directed edge $(u,v)$ appearing consecutively in $P_{\text{fast}}$}{
        $p_{ed}(u,v) \leftarrow \max\{p_{ed}(u,v),0.8\}$ \;
    }
    
    \tcc{Stage 3: full BipartiteGBS-guided genetic search}
    $result \leftarrow \textsc{BipartiteGBSGAEvolutionLoop}(G,p_{ed},\mathcal{S},N_{\text{pop}},G_{\max}/2,p_c,p_m,\alpha_{\max},\beta)$ (Alg.~\ref{alg:gbs_ga_main}) \;
    
    \Return $result$ \;
}
\end{algorithm}

\appsection{Details for the classical benchmark algorithms}\label{appx:classical-benchmarks}

This appendix collects the three classical benchmark algorithms used in the numerical comparison. These baselines isolate the contribution of classical greedy search, unguided evolutionary search, and simple degree-based structural guidance from the BipartiteGBS-enhanced components. The pseudocode below is written to match the implementation used for the reported figures, including the corrected Hamiltonian-cycle validation rule in Alg.~\ref{alg:gbs_helper}.

\appsubsection{Nearest neighbor heuristic}\label{appx:nn-baseline}

The nearest neighbor heuristic (NN) is a greedy baseline for directed graphs. It randomly shuffles the starting vertices and attempts up to $\min(10,n)$ of them. From each start, it repeatedly appends the smallest-indexed unvisited outgoing neighbor. If no admissible outgoing edge remains, the current path terminates. The algorithm returns a Hamiltonian cycle as soon as one is found; otherwise, it reports the longest path encountered across all attempted starting vertices.

\begin{algorithm}[H]
\caption{Nearest Neighbor Heuristic (NN)}
\label{alg:nn_baseline}
\DontPrintSemicolon
\SetKwInOut{Input}{Input}
\SetKwInOut{Output}{Output}

\Input{
    Directed graph $G=(V,E)$, $|V|=n$
}
\Output{
    Hamiltonian cycle if found; otherwise the longest path found
}

\Begin{
    $best \leftarrow []$ \;
    $\text{starts} \leftarrow \text{random shuffle of } [0,\ldots,n-1]$ \;

    \ForEach{$s \in \text{starts}[0:\min(10,n)]$}{
        $\text{path} \leftarrow [s]$, $\text{used} \leftarrow \{s\}$, $\text{current} \leftarrow s$ \;

        \While{$|\text{path}| < n$}{
            $\text{neighbors} \leftarrow \{v : G.\text{has\_edge}(\text{current}, v) \text{ and } v \notin \text{used}\}$ \;
            \If{$\text{neighbors} = \emptyset$}{\textbf{break}}
            $\text{next} \leftarrow \min(\text{neighbors})$ \;
            $\text{path}.\text{append}(\text{next})$, $\text{used}.\text{add}(\text{next})$ \;
            $\text{current} \leftarrow \text{next}$ \;
        }

        \If{$|\text{path}| > |best|$}{$best \leftarrow \text{path}$}
        \If{$|\text{path}|=n$ and $G.\text{has\_edge}(\text{path}[-1], \text{path}[0])$}{\Return $\text{path}$}
    }

    \Return $best$ \;
}
\end{algorithm}

\appsubsection{Standard genetic algorithm}\label{appx:standard-ga}

The standard genetic algorithm (GA) is the unguided evolutionary baseline. It uses the same permutation representation as \BipartiteGBSGA{} but evaluates individuals only by the fraction of valid directed edges in the circular permutation. Selection and order crossover reuse the helper routines in Alg.~\ref{alg:gbs_helper}, while mutation is implemented by swapping two randomly selected vertices.

\begin{algorithm}[H]
\caption{Standard Genetic Algorithm (GA)}
\label{alg:standard_ga}
\DontPrintSemicolon
\SetKwInOut{Input}{Input}
\SetKwInOut{Output}{Output}

\Input{
    Directed graph $G=(V,E)$, $|V|=n$ \\
    Population size $N_{\text{pop}}$, maximum generations $G_{\max}$ \\
    Crossover rate $p_c$, mutation rate $p_m$
}
\Output{
    Hamiltonian cycle if found; otherwise the longest path found
}

\Begin{
    $P \leftarrow \{\text{random permutations of } V\text{ of size }N_{\text{pop}}\}$ \;
    $best \leftarrow \text{NULL}$, $bestFit \leftarrow -1$ \;
    
    \For{$gen \leftarrow 1$ \KwTo $G_{\max}$}{
        \ForEach{$ind \in P$}{
            $fitness[ind] \leftarrow \text{number of valid edges in the circular permutation } ind \,/\, n$ \;
        }
        $best_{\text{gen}} \leftarrow \arg\max_{ind \in P} fitness[ind]$ \;
        \If{$fitness[best_{\text{gen}}] > bestFit$}{
            $best \leftarrow best_{\text{gen}}$, $bestFit \leftarrow fitness[best_{\text{gen}}]$ \;
        }
        \If{$\textsc{IsHamiltonianCycle}(best,G)$ (Alg.~\ref{alg:gbs_helper})}{\Return $best$}
        
        $P_{\text{new}} \leftarrow \emptyset$ \;
        \For{$i \leftarrow 1$ \KwTo $N_{\text{pop}}$}{
            $p_1 \leftarrow \textsc{TournamentSelect}(P, fitness)$ (Alg.~\ref{alg:gbs_helper}) \;
            $p_2 \leftarrow \textsc{TournamentSelect}(P, fitness)$ (Alg.~\ref{alg:gbs_helper}) \;
            \uIf{$\text{random}() < p_c$}{
                $child \leftarrow \textsc{OrderCrossover}(p_1,p_2)$ (Alg.~\ref{alg:gbs_helper}) \;
            }
            \Else{
                $child \leftarrow p_1$ \;
            }
            \If{$\text{random}() < p_m$}{
                choose two distinct indices $i,j$ uniformly at random and swap $child[i]$ with $child[j]$ \;
            }
            $P_{\text{new}}.\text{append}(child)$ \;
        }
        $P \leftarrow P_{\text{new}}$ \;
    }
    
    \Return longest path from $best$ \;
}
\end{algorithm}

\appsubsection{Degree-centrality-enhanced genetic algorithm}\label{appx:degree-ga}

The degree-centrality-enhanced genetic algorithm (Degree-GA) is a classical structural baseline. Instead of using BipartiteGBS-derived co-occurrence frequencies, it assigns each directed edge $(u,v)$ a score based on the out-degree of $u$ and the in-degree of $v$, and then uses this score to bias initialization, fitness evaluation, and mutation. This comparison tests whether simple vertex-level degree information can explain the advantage of the BipartiteGBS-guided search. For readability, Alg.~\ref{alg:degree_ga} gives the main evolutionary loop, while Algs.~\ref{alg:degree_mutate} and~\ref{alg:degree_greedy_construct} give the two helper routines.

\begin{algorithm}[H]
\caption{DegreeCentrality-Enhanced Genetic Algorithm (Degree-GA)}
\label{alg:degree_ga}
\DontPrintSemicolon
\SetKwInOut{Input}{Input}
\SetKwInOut{Output}{Output}

\Input{
    Directed graph $G=(V,E)$, $|V|=n$ \\
    Population size $N_{\text{pop}}$, max generations $G_{\max}$, \\
    Crossover rate $p_c$, mutation rate $p_m$, weight $\alpha$
}
\Output{
    Best found Hamiltonian cycle or longest path
}

\Begin{
    \tcc{Compute degree information}
    \ForEach{$v \in V$}{
        $\deg_{\text{out}}(v) \leftarrow |\{u: (v,u)\in E\}|$ \;
        $\deg_{\text{in}}(v) \leftarrow |\{u: (u,v)\in E\}|$ \;
    }
    $d_{\text{total}}(v) \leftarrow \deg_{\text{out}}(v) + \deg_{\text{in}}(v)$ for all $v \in V$ \;
    \ForEach{$(u,v)\in E$}{
        $s(u,v) \leftarrow \frac{1}{2}\bigl(\frac{\deg_{\text{out}}(u)}{\max\deg_{\text{out}}} + \frac{\deg_{\text{in}}(v)}{\max\deg_{\text{in}}}\bigr)$ \;
    }

    \tcc{Population initialization (half random, half degree-greedy)}
    $\text{pop} \leftarrow \emptyset$ \;
    \For{$i=1$ \KwTo $\lfloor N_{\text{pop}}/2\rfloor$}{$\text{pop}.\text{append}(\text{random permutation})$}
    \For{$i=1$ \KwTo $\lfloor N_{\text{pop}}/2\rfloor$}{
        $\text{pop}.\text{append}(\textsc{GreedyConstruct}(G, s, d_{\text{total}}))$ (Alg.~\ref{alg:degree_greedy_construct}) \;
    }
    \While{$|\text{pop}| < N_{\text{pop}}$}{$\text{pop}.\text{append}(\text{random permutation})$}

    \tcc{Evolution}
    \For{$gen=1$ \KwTo $G_{\max}$}{
        $\alpha_{\text{cur}} \leftarrow \alpha \cdot gen/G_{\max}$ \;
        \ForEach{$ind \in \text{pop}$}{
            $len \leftarrow \text{number of valid consecutive edges in } ind \text{ (including the closing edge if it exists)}$ \;
            $\text{fit}[ind] \leftarrow (1-\alpha_{\text{cur}})\cdot\frac{len}{n} + \alpha_{\text{cur}}\cdot\frac{\sum_{i=0}^{n-1} s(ind[i],ind[(i+1)\bmod n])}{n}$ \;
            \tcc{Use $s(u,v)=0$ for nonexistent or undefined edges in the summation}
        }
        $best \leftarrow \arg\max \text{fit}$ \;
        \If{$best$ is Hamiltonian cycle}{\Return $best$}

        $\text{new\_pop} \leftarrow \emptyset$ \;
        \For{$i=1$ \KwTo $N_{\text{pop}}$}{
            $p_1 \leftarrow \textsc{TournamentSelect}(\text{pop},\text{fit})$ (Alg.~\ref{alg:gbs_helper}) \;
            $p_2 \leftarrow \textsc{TournamentSelect}(\text{pop},\text{fit})$ (Alg.~\ref{alg:gbs_helper}) \;
            $child \leftarrow \textsc{OrderCrossover}(p_1,p_2)$ (Alg.~\ref{alg:gbs_helper}) if random $<p_c$ else $p_1$ \;
            \If{random $<p_m$}{
                $child \leftarrow \textsc{DegreeMutate}(child, s)$ (Alg.~\ref{alg:degree_mutate}) \;
            }
            $\text{new\_pop}.\text{append}(child)$ \;
        }
        $\text{pop} \leftarrow \text{new\_pop}$ \;
    }
    \Return longest path from best \;
}
\end{algorithm}

\begin{algorithm}[H]
\caption{Degree-Guided Mutation}
\label{alg:degree_mutate}
\DontPrintSemicolon
\SetKwInOut{Input}{Input}
\SetKwInOut{Output}{Output}

\Input{
    Individual $ind$, edge scores $s$
}
\Output{
    Mutated individual
}

\Begin{
    $n \leftarrow |ind|$ \;
    \If{$\text{random}() < 0.7$}{
        \For{$i=0$ \KwTo $n-1$}{
            $score_i \leftarrow s(ind[i], ind[(i+1)\bmod n])$ (default $0.5$ if undefined) \;
        }
        $idx \leftarrow \arg\min_i score_i$ \;
        $jdx \leftarrow \text{random}(0, n-1)$ \;
        $\text{swap}(ind[idx], ind[jdx])$ \;
        \Return $ind$ \;
    }
    $(i,j) \leftarrow \text{random sample of size } 2 \text{ from } [0,\dots,n-1]$ \;
    $\text{swap}(ind[i], ind[j])$ \;
    \Return $ind$ \;
}
\end{algorithm}

\begin{algorithm}[H]
\caption{Degree-Greedy Construction}
\label{alg:degree_greedy_construct}
\DontPrintSemicolon
\SetKwInOut{Input}{Input}
\SetKwInOut{Output}{Output}

\Input{
    Directed graph $G=(V,E)$, edge scores $s$, total degree scores $d_{\text{total}}$
}
\Output{
    A vertex permutation
}

\Begin{
    $n \leftarrow |V|$, $\text{used} \leftarrow \emptyset$, $\text{path} \leftarrow []$ \;

    \tcc{Start with vertex of highest total degree}
    $\text{start} \leftarrow \arg\max_{v \in V} d_{\text{total}}(v)$ \;
    $\text{path}.\text{append}(\text{start})$, $\text{used}.\text{add}(\text{start})$ \;

    \While{$|\text{path}| < n$}{
        $\text{current} \leftarrow \text{path}[-1]$ \;
        $\text{neighbors} \leftarrow \{v: G.\text{has\_edge}(\text{current}, v) \text{ and } v \notin \text{used}\}$ \;

        \If{$\text{neighbors} \neq \emptyset$}{
            $\text{next} \leftarrow \arg\max_{v \in \text{neighbors}} s(\text{current}, v)$ \tcc*[r]{use edge score, not total degree}
            $\text{path}.\text{append}(\text{next})$, $\text{used}.\text{add}(\text{next})$ \;
        }
        \Else{
            $\text{remaining} \leftarrow V \setminus \text{used}$ \;
            $\text{path}.\text{append}(\text{random.choice}(\text{remaining}))$ \;
            $\text{used}.\text{add}(\text{path}[-1])$ \;
        }
    }
    \Return $\text{path}$ \;
}

\end{algorithm}

\end{document}